\documentclass[aps,pra,twocolumn,groupedaddress,showpacs]{revtex4-1}
\usepackage{graphicx}
\usepackage{amsmath}
\usepackage{amssymb}
\usepackage{epstopdf}
\usepackage{pstool}
\usepackage{float}
\usepackage{pstricks}           
\usepackage{dcolumn}            
\usepackage{bm}                 
\usepackage{hyperref}           
\usepackage{hypernat}           
\usepackage[latin1]{inputenc}   
\usepackage[T1]{fontenc}
\usepackage[english]{babel}
\usepackage{soul}
\usepackage{rotating}
\usepackage{subfigure}
\usepackage{slashed}
\usepackage{color} 

\newcommand{\bk}{\textbf{k}}
\newcommand{\bp}{\textbf{p}}
\newcommand{\bv}{\textbf{v}}
\newcommand{\br}{\textbf{r}}

\newcommand{\bE}{\textbf{E}}
\newcommand{\bB}{\textbf{B}}

\newcommand{\bA}{\textbf{A}}

\newcommand*{\brt}{(\br,t)}
\newcommand*{\brtp}{(\br,t')}
\newcommand*{\tbr}{\tilde{\br}}

\newcommand{\zu}{\hat{\textbf{z}}}

\newcommand{\pt}{\frac{\partial}{\partial t}}
\newcommand{\dt}{\frac{d}{dt}}

\newcommand*{\re}[1]{\text{Eq. (}\ref{#1}\text{)}}
\newcommand*{\res}[1]{\text{Eqs. (}\ref{#1}\text{)}}
\newcommand*{\rf}[1]{\text{Fig. }\ref{#1}}

\newcommand*{\frf}[1]{\text{Figure }\ref{#1}}
\newcommand{\Ef}{\mathcal{E}_f}

\begin{document}

\title{A semi-classical approach for solving the time-dependent
Schr\"{o}dinger equation in spatially inhomogeneous electromagnetic pulses}

\author{Jianxiong Li and Uwe Thumm}

\affiliation{Department of Physics, Kansas State University,
Manhattan, Kansas 66506, USA }
\date{\today}

\begin{abstract}
To solve the time-dependent Schr\"{o}dinger equation in spatially
inhomogeneous pulses of electromagnetic radiation, we propose an
iterative semi-classical complex trajectory approach. In numerical
applications, we validate this method against {\it ab initio}
numerical solutions by scrutinizing (a) electronic states in
combined Coulomb and spatially homogeneous laser fields and (b)
streaked photoemission from hydrogen atoms and plasmonic gold
nanospheres. In comparison with streaked photoemission calculations
performed in strong-field approximation, we demonstrate the improved
reconstruction of the spatially inhomogeneous induced plasmonic
infrared field near a nanoparticle surface from streaked
photoemission spectra.
\end{abstract}


\maketitle

\section{Introduction} \label{sec:intro}

The exposure of gaseous atomic, mesoscopic, and solid targets to
incident pulses of electromagnetic radiation of sufficiently high
photon energy or intensity leads to the emission of
photoelectrons~\cite{Hufner2003}. For more than a century,
photoelectron spectroscopy has very successfully exploited this
phenomenon and has long become established as one of the most
prolific techniques for unraveling the {\em static} electronic
structure of matter by examining the kinetic-energy or momentum
distribution of emitted photoelectrons. More recently, starting in
the 21st century, advances in ultrafast laser technology started to
extend photoemission spectroscopy into the time
domain~\cite{Hentschel2001,Chang2004,Sansone2006}. Importantly, the
development of attosecond streaking~\cite{Krausz2009,Thumm2015Chap}
and interferometric~\cite{Paul2001,Klunder2011,Locher2015}
photoelectron spectroscopy enabled the observation of electron
dynamics at the natural time scale of the electron motion in matter
(attoseconds, $1$~as $= 10^{-18}$~s). This was demonstrated in
proof-of-principle experiments for gaseous
atomic~\cite{Drescher2002,Kienberger2004,Johnsson2007,Wang2010,Schultze2010,Ott2013,Bernhard2014}
and molecular~\cite{Niikura2003,Kling2006,Staudte2007} targets.
Attosecond time-resolved photoemission spectroscopy is currently
being extended to complex targets~\cite{Leone2014,Thumm2015Chap},
such as nanostructures and
nanoparticles~\cite{Forg2016,Li2016,Seiffert2017,Schoetz2017,Saydanzad2017,Li2017,Saydanzad2018,Li2018PRL},
and solid
surfaces~\cite{Locher2015,Lucchini2015,Chen2017,Neppl2015,Tao2016,Heinzmann2017,Kasmi2017,Ambrosio2018,Ambrosio2019},
making it possible to examine, for example, the dynamics of
photoemission from a surface  on an absolute time
scale~\cite{Ossiander2018} and suggesting, for example, the
time-resolved observation of the collective motion of electrons
(plasmons) in condensed-matter
systems~\cite{Zhang2011,Chew2012,Lupetti2014}.

In combination with advances in nanotechnology, allowing the
production of plasmonic nanostructures with increasing efficiency at
the nm length scale, attosecond photoemission spectroscopy has
started to progress towards the spatiotemporal imaging of electron
dynamics in complex targets, approaching the atomic length and time
scales (nm and
attoseconds)~\cite{Lemke2013,Leone2014,Liao2015,Neppl2015,Forg2016,Saydanzad2017,Li2017,Saydanzad2018,Li2018PRL}.
Photo\-emission spectroscopy therefore holds promise to become a
powerful tool for examining nm-attosecond scale processes that are
operative in plasmonically enhanced
photocatalysis~\cite{Schlather2017}, light
harvesting~\cite{Sheldon2014}, surface-enhanced Raman
spectroscopy~\cite{Le2008}, biomedical and chemical
sensing~\cite{Kabashin2009}, tumor detection and
treatment~\cite{Ayala2014}, and ultrafast electro-optical
switching~\cite{Krausz2014}. The concurrent development and
provision of large-scale light sources, capable of producing intense
ultrashort pulses in the extreme ultraviolet (XUV) to X-ray spectral
range at several leading laboratories in Europe, the United Stated,
and Japan~\cite{NatFELcollect2019}, promises to further boost the
value of spatiotemporally resolved electron spectroscopy as a tool
for imaging electronic dynamics within a wide array of basic and
applied research projects.

Being able to take advantage of the full potential offered by
current and emerging atomic scale photoelectron imaging techniques
relies on theoretical and numerical modeling. This is true for
comparatively simple atoms in the gas phase, and for complex
nanostructured targets additional theoretical challenges
arise~\cite{Leone2014,Thumm2015Chap}. While for atomic
photoionization by visible and near UV light, the size of the target
is small compared to the wavelength of the incident light pulse,
this is no longer true for X-ray ionization, leading to the
well-know breakdown of the dipole
approximation~\cite{Merzbacher1998}. Furthermore, for
nanoparticles~\cite{Seiffert2017,Schoetz2017,Saydanzad2017,Li2016,Li2017,Saydanzad2018,Li2018PRL},
(nanostructured)
surfaces~\cite{Obreshkov2006,Liao2014PRL,Liao2014,Ambrosio2019}, and
layered
structures~\cite{Liao2015,Neppl2015Nature,Ambrosio2018,Ambrosio2019},
not only the comparability of the wavelength and structure size
requires careful quantum-mechanical modeling beyond the dipole
approximation, but also the target's spatially inhomogeneous
dielectric response to the incident light
pulse~\cite{Liao2014PRL,Liao2014}. Most numerical models for
streaked and interferometric photoemission from atoms are based on
the so-called `strong-field approximation
(SFA)'~\cite{Thumm2015Chap}. The SFA builds on the assumption that
photo-emitted electrons are solely exposed to {\em spatially
homogeneous} external fields. It discards all other interactions
photo-released electrons may be subject to (e.g., with the residual
parent ion) and cannot accommodate spatially inhomogeneous
final-state interactions.

While the SFA was shown to deteriorate for lower photoelectron
energies~\cite{Zhang2010}, it completely looses its applicability
for complex targets as screening and plasmonic effects expose
photoelectrons to {\em inhomogeneous} net electro-magnetic
fields~\cite{Zhang2011,Liao2014PRL,Liao2014,Thumm2015Chap}.  The
convenient use of analytically known so-called `Volkov
wavefunctions' for the photoelectron's motion in homogeneous
electromagnetic fields~\cite{Volkov1935} is no longer acceptable,
since dielectric response effects entail screening length and
induced plasmonic fields at the nm length
scale~\cite{Zhang2011,Schoetz2017,Saydanzad2017,Li2016,Li2017,Saydanzad2018,Li2018PRL,Ambrosio2019}.
Thus, the numerical modeling of photoemission from complex targets
with morphologies or plasmonic response lengths at the nm scale by
intense short wavelength pulses (made increasingly available at new
(X)FEL light sources~\cite{NatFELcollect2019}), necessitates
photoemission models beyond the SFA.

To this effect we previously employed heuristically generalized
Volkov states to model photoemission from bare and adsorbate-covered
metal surfaces~\cite{Liao2014PRL,Liao2014,Ambrosio2018,Ambrosio2019}
and plasmonic nanoparticles~\cite{Li2016,Li2017,Li2018PRL}. While
this allowed us to numerically model
streaked~\cite{Zhang2009PRL,Liao2014PRL,Liao2014,Liao2015} and
interferometric photoemission spectra from
surfaces~\cite{Ambrosio2018,Ambrosio2019}, in fair to good agreement
with experimental data, and to reconstruct plasmonic fields near
gold nanospheres~\cite{Li2018PRL}, a systematic mathematical
solution of the time-dependent Schr\"{o}dinger (TDSE) for a single
active electron exposed to inhomogeneous external fields remains to
be explored. We here discuss a semiclassical model for obtaining
such solutions. While being approximate, our complex-phase
Wentzel-Kramer-Brillouin (WKB)-type approach  lends itself to
systematic iterative refinement. Our proposed method, termed ACCTIVE
(Action Calculation by Classical Trajectory Integration in Varying
Electromagnetic fields), employs complex classical trajectories to
solve the TDSE in the presence of spatially inhomogeneous
electromagnetic pulses that are represented by time-dependent
inhomogeneous scalar and vector potentials. Our approach is inspired
by the semiclassical complex-trajectory method for solving the TDSE
with time-independent scalar interactions of Boiron and
Lombardi~\cite{Boiron1998} and its adaptation to time-dependent
scalar interactions by Goldfarb, Schiff, and
Tannor~\cite{Goldfarb2008}.

Following the mathematical formulation of ACCTIVE in
Sec.~\ref{sec:theory}, we validate this method by discussing five
examples in Sec.~\ref{sec:examples}. We first compare  ACCTIVE
calculations with {\it ab initio} numerical solutions by
scrutinizing electronic states in a (i) homogeneous laser field,
(ii) Coulomb field, and (iii) combination of laser and Coulomb
fields. Next, we apply ACCTIVE to streaked photoemission from (iv)
hydrogen atoms and (v) plasmonic nanoparticles. In the application
to Au nanospheres, we examine final states for the simultaneous
interaction of the photoelectron with the spatially inhomogeneous
plasmonically enhanced field induced by the streaking infrared (IR)
laser pulse and demonstrate the improved reconstruction of the
induced nanoplasmonic IR field from streaked photoemission spectra.
Section~\ref{sec:conclusion} contains our summary. In three
appendices we add details of our calculations  within ACCTIVE of
Volkov wavefunctions (Appendix~\ref{app:A}) and Coulomb
wavefunctions (Appendix~\ref{app:B}), and additional comments on
streaked photoemission from Au nanospheres (Appendix~\ref{app:C})
within ACCTIVE.

\section{Theory} \label{sec:theory}

We seek approximate solutions of the TDSE for a particle of
(effective) mass $m$ and charge $q$ in an inhomogeneous
time-dependent electro-magnetic field given by the scalar and vector
potentials $\phi\brt$ and $\bA\brt$ and an additional scalar
potential $V\brt$,
\begin{equation}\label{eq:TDSE}
 i\hbar\pt \Psi\brt = \Bigg\{ \frac{1}{2m} \Big[ i\hbar\nabla + q\bA\brt \Big]^2 + \varphi\brt \Bigg\}
 \Psi\brt,
\end{equation}
where $\varphi\brt = q\phi\brt + V\brt$ and $V\brt$ is any scalar
potential. Representing the wavefunction in eikonal form, $\Psi\brt
= e^{iS\brt/\hbar}$, Eq.~(\ref{eq:TDSE}) can be rewritten in terms
of the complex-valued quantum-mechanical action $S\brt$,
\begin{align}
 \pt S\brt
    &+ \frac{1}{2m}\Big[\nabla S\brt - q\bA\brt \Big]^2 + \varphi\brt \nonumber \\
    &= \frac{i\hbar}{2m}\nabla \cdot \Big[\nabla S\brt -q\bA\brt \Big].\label{eq:STDSE}
\end{align}
Expanding the action in powers of $\hbar$
\cite{Boiron1998,Goldfarb2008},
\begin{equation}\label{eq:SExpansion}
 S\brt = \sum_{n=0}^{\infty} \hbar^n S_n\brt,
\end{equation}
substituting \re{eq:SExpansion} into \re{eq:STDSE}, and comparing
terms of equal order, results in the set of coupled partial
differential equations
\begin{subequations}\label{eq:SPDEs}
 \begin{eqnarray}
  \pt S_0\brt
    &+& \frac{\big[\nabla S_0\brt - q\bA\brt \big]^2}{2m} + \varphi\brt = 0  \quad\quad \label{eq:S0PDE} \\
  \pt S_1\brt
    &+& \bigg[ \frac{\nabla S_0\brt - q\bA\brt}{m} \bigg] \cdot \nabla S_1\brt \nonumber\\
    &=& \frac{i}{2}\nabla \cdot \bigg[ \frac{\nabla S_0\brt - q\bA\brt}{m} \bigg] \label{eq:S1PDE}\\
  \pt S_n\brt
    &+& \bigg[ \frac{\nabla S_0\brt - q\bA\brt}{m} \bigg] \cdot \nabla S_n\brt \nonumber\\
    &=& - \frac{1}{2m}\sum_{j=1}^{n-1}  \nabla S_j\brt \cdot \nabla S_{n-j}\brt \nonumber\\
    &\quad&\quad\quad + \frac{i}{2m}\nabla^2S_{n-1}\brt \quad\quad (n\geq2), \label{eq:SnPDE}
 \end{eqnarray}
\end{subequations}
where the lowest-order contribution $S_0\brt$ is the classical
action of a charged particle moving in the electromagnetic field
given by $\bE\brt = - \nabla \varphi\brt/q - \partial
\bA\brt/\partial t$ and $\bB\brt = \nabla\times\bA\brt$.

Solving the classical Hamilton-Jacobi equation (HJE) \re{eq:S0PDE}
leads to  Newton's Second Law,
\begin{equation}\label{eq:Newton2}
\dt\bv\brt = \frac{q}{m} \Big[ \bE\brt + \bv\brt\times\bB\brt
\Big],
\end{equation}
where the classical velocity field $\bv\brt$ and kinetic momentum,
\begin{equation} \label{eq:vField}
\bp\brt \equiv m\bv\brt \equiv \nabla S_0\brt - q\bA\brt,
\end{equation}
are given in terms of the canonical momentum $\nabla
S_0\brt$~\cite{Goldstein}. The combination of the HJE
(\ref{eq:S0PDE}) and Eq.~(\ref{eq:vField}) provides the Lagrangian
$L\big[\br,\bv\brt,t\big]$ as a total time differential of
$S_0\brt$,
\begin{align}
\dt S_0
&\brt = L\big[\br,\bv\brt,t\big] \nonumber\\
&= \frac{1}{2}m\bv^2\brt + q\bv\brt\cdot\bA\brt -\varphi\brt.
\label{eq:S0TTD}
\end{align}

Similarly, by substituting \re{eq:vField} into Eqs.~(\ref{eq:S1PDE})
and (\ref{eq:SnTTD}), we find the total time derivatives of the
first-order contribution to $S\brt$,
\begin{equation}
\dt S_1\brt = \frac{i}{2}\nabla\cdot\bv\brt, \label{eq:S1TTD}
\end{equation}
and of all higher order terms,
\begin{align}
\dt S_n\brt
&= - \frac{1}{2m}\sum_{j=1}^{n-1}  \nabla S_j\brt \cdot \nabla S_{n-j}\brt\nonumber\\
&\quad\quad\quad + \frac{i}{2m}\nabla^2S_{n-1}\brt  \quad\quad
(n\geq2). \label{eq:SnTTD}
\end{align}
Approximate solutions to $S\brt$ can be obtained by iteration of
\re{eq:SnTTD}, after integrating the total time derivatives in
\res{eq:S0TTD}, (\ref{eq:S1TTD}), and (\ref{eq:SnTTD}) along
classical trajectories $\tilde{\br}(t)$ that are defined by
\begin{equation} \label{eq:vDef}
\dt\tilde{\br}(t) \equiv \bv\big[\tilde{\br}(t),t\big]
\end{equation}
with respect to a reference time (integration constant) $t_r$. The
wavefunction at $t_r$, $\Psi_r(\br) = \Psi(\br,t_r)$, provides
initial ($t_r \ll 0$) or asymptotic ($t_r \gg 0$) conditions in
terms of the action
\begin{equation}\label{eq:ICS}
 S(\br,t_r) = -i\hbar\ln[\Psi_r(\br)]
\end{equation}
and the velocity field
\begin{align}
 \bv(\br,t_r)
   &= -\frac{1}{m}\nabla S_0(\br,t_r) - \frac{q}{m}\bA(\br,t_r) \nonumber \\
   &\approx -\frac{1}{m}\nabla S(\br,t_r) - \frac{q}{m}\bA(\br,t_r) \nonumber \\
   &= -\frac{i\hbar\nabla\Psi_r(\br)}{m\Psi_r(\br)} - \frac{q}{m}\bA(\br,t_r). \label{eq:IC}
\end{align}

\begin{figure}
    \includegraphics[width=1.00\linewidth]{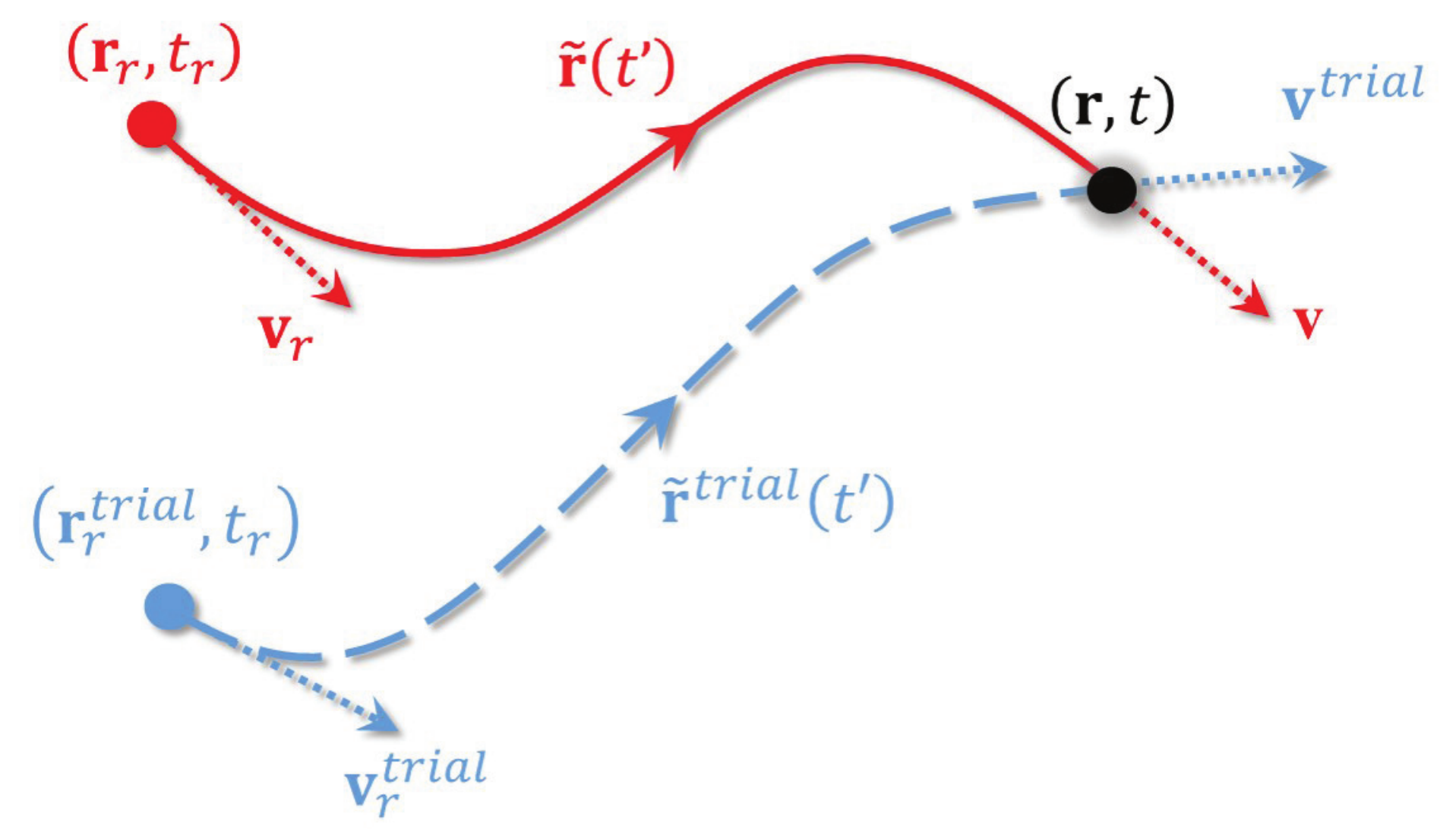}
\caption{(Color online) Illustration of the shooting method used for
determining classical trajectories. For any given event $(\br,t)$
and a predetermined reference time $t_r$, trajectories are
classically propagated from trial points in phase space,
$(\br,\bv^{trial})$, at time $t$ along trial trajectories
$\tilde{\br}^{trial}(t')$. The velocity field $\bv$ and appropriate
trajectory $\tilde{\br}(t')$ are determined by iterating the trial
velocity $\bv^{trial}$, in order to find the roots of
$f(\bv^{trial})$ in Eq.~(\ref{eq:shooting}). } \label{fig:1}
\end{figure}

The semiclassical solution of \res{eq:S0TTD}, (\ref{eq:S1TTD}), and
(\ref{eq:SnTTD}) requires an appropriate classical trajectory
$\tbr(t')$ - for any given `current' event $(\br,t)$ - that connects
the `current' coordinate and velocity,
\begin{equation} \label{eqs:CTCurrent}
 \br=\tbr(t),  \quad\quad\quad  \bv=\frac{d\tilde{\br}(t')}{dt'}\bigg|_t,
\end{equation}
to the proper coordinate and velocity at $t_r$,
\begin{subequations} \label{eqs:CTInitial}
    \begin{eqnarray}
    \br_r &=& \tilde{\br}(t_r), \label{eq:CTInitial1}\\
    \bv_r &=& \frac{d\tilde{\br}(t')}{dt'}_{t_r}=-\frac{i\hbar\nabla\Psi_r(\br_r)}{m\Psi_r(\br_r)} - \frac{q}{m}\bA(\br_r,t_r).\quad \label{eq:CTInitial2}
    \end{eqnarray}
\end{subequations}
The known quantities in \res{eqs:CTCurrent} and
(\ref{eqs:CTInitial}) are $\br$, $t$, and $t_r$, while $\bv$,
$\br_r$, and $\bv_r$ are to be determined. To numerically calculate
the undetermined quantities, we employ a shooting method, starting
with a `trial' velocity $\bv^{trial}$ at position $\br$ and time
$t$. Propagating $\br$ to the reference time according to
\re{eq:Newton2} results in $\br_r^{trial} = \tbr^{trial}(t_r)$ and
$\bv_r^{trial} =d\tilde{\br}^{trial}(t') /dt' \mid_{t_r}$
(Fig.~\ref{fig:1}).

The velocity field $\bv$ that satisfies \re{eq:Newton2} can now be
found numerically by minimizing the function
\begin{equation} \label{eq:shooting}
  f(\bv^{trial}) = \mid \bv^{trial}_r + \frac{i\hbar\nabla\Psi_r(\br^{trial}_r)}{m\Psi_r(\br^{trial}_r)}
  + \frac{q}{m}\bA(\br^{trial}_r,t_r) \mid
\end{equation}
for an appropriate range of start trial velocities. In our numerical
applications this is accomplished by an efficient multi-dimensional
quasi-Newton root-finding algorithm (Broyden's
method)~\cite{Broyden,Atkinson}. Once the correct trajectories
$\tbr(t')$ are determined by finding the roots of
Eq.~(\ref{eq:shooting}), the actions in Eqs.~(\ref{eq:S0TTD}),
(\ref{eq:S1TTD}), and (\ref{eq:SnTTD}) are integrated along these
trajectories and composed - by truncating Eq.~(\ref{eq:SExpansion})
- into an approximate solution of Eq.~({\ref{eq:TDSE}).

Since each term $S_n\brt$ in Eq.~(\ref{eq:SExpansion}) depends only
on terms of lower orders, ACCTIVE enables, in principle, the
systematic iterative refinement of approximate solutions of
Eq.~(\ref{eq:TDSE}) by including successively higher orders $n$. The
iteration is started with $S_0\brt$, which is determined by the
velocity field $\bv\brt$, and continued by integrating
Eqs.~(\ref{eq:S1TTD}) and (\ref{eq:SnTTD}).


In the numerical examples discussed in Sec.~\ref{sec:examples}
below, we find that retaining only the zero'th and first-order
terms, $S_0\brt$ and $S_1\brt$, provides sufficiently accurate and
physically meaningful solutions at modest numerical expense. Thus,
according to Eqs.~(\ref{eq:S0TTD}) and (\ref{eq:S1TTD}), we apply
\begin{align}
 \Psi\brt &
    \approx \exp \big\{iS_0\brt/\hbar + iS_1\brt\big\} \nonumber\\
    &= e^{iS(\br_r,t_r)/\hbar} \exp \Bigg\{-\frac{1}{2}\int_{t_r}^{t}\nabla\cdot\bv\Big(\tbr(t'),t'\Big) dt' \nonumber\\
    &\quad\quad +\frac{i}{\hbar}\int_{t_r}^{t} L\bigg[\tbr(t'),\bv\Big(\tbr(t'),t'\Big),t'\bigg] dt'\Bigg\}. \label{eq:GV}
\end{align}
For real classical trajectories and potentials, the integral of
$S_0(\br,t)$ is real, representing a local phase factor, while
$S_1(\br,t)$ is purely imaginary and defines the wavefunction
amplitude, as in the standard WKB approach \cite{Merzbacher1998}.
The quantum-mechanical probability density $\rho\brt$ then satisfies
the continuity equation,
\begin{equation}
 \frac{d\rho\brt}{dt} = \dt|\Psi\brt|^2 = -\rho\brt\nabla\cdot\bv\brt,
\end{equation}
for the classical probability flux
$\rho\brt\,\bv\brt$~\cite{Batchelor}.

\section{Examples} \label{sec:examples}

We validate the ACCTIVE method by discussing five applications to
electron wavefunctions in Coulomb and laser fields.

\subsection{Volkov wavefunction} \label{sec:Volkov}

For the simple example of an electron in a time-dependent, spatially
homogeneous laser field, the potentials in Eq.~(\ref{eq:TDSE}) and
reference wavefunction are (in the Coulomb electromagnetic
gauge~\cite{Merzbacher1998})
\begin{equation} \label{eq:ICV}
  \bA\brt = \bA(t) ,\quad \varphi\brt = 0 ,\quad \Psi_r(\br) =
  e^{i\bp\cdot\br/\hbar},
\end{equation}
and the first-order wavefunction in Eq.~(\ref{eq:GV}) reproduces the
well-known analytical Volkov solution~\cite{Volkov1935},
\begin{equation}
  \Psi^V\brt = \exp
    \bigg\{
      \frac{i~\bp\cdot\br}{\hbar}
      -\frac{i}{2m\hbar}\int_{t_r}^t
        \big[
          \bp - q\bA(t')
        \big]^2 dt'
    \bigg\}. ~~\label{eq:Volkov}
\end{equation}
For details of the derivation of \re{eq:Volkov} within ACCTIVE see
Appendix~\ref{app:A}.

\subsection{Coulomb wavefunction} \label{sec:Coulomb}

As a second simple example and limiting case, we consider an unbound
electron in the Coulomb field of a proton. In this case the
potentials in Eq.~(\ref{eq:TDSE}) are
\begin{equation}
  \bA\brt = 0 ,\quad \varphi\brt = -k_e\frac{e^2}{r},
\end{equation}
where $e$ is the elementary charge and $k_e$ the electrostatic
constant. Assuming outgoing-wave boundary conditions, we define the
reference wavefunction at a sufficiently large reference time $t_r$
as the `outgoing' Coulomb wave
\begin{equation}\label{eq:ICCW}
  \Psi_r(\br,t_r) \xrightarrow{~t_r\rightarrow\infty,~z\rightarrow+\infty~}
  e^{i\left(kz-\frac{\hbar
  k^2}{2m}t_r  \right)}.
\end{equation}
Here $\br=(x,y,z)$ and $p=\hbar k>0$ is the final electron momentum.
In this case the TDSE is solved exactly by the well-known Coulomb
wavefunction
\begin{equation} \label{eq:CoulombAna}
  \Psi^C_k(\br,t) = \frac{e^{\frac{\pi}{2k}}\Gamma(1-i/k)}{(2\pi)^{3/2}} {}_1F_1(i/k,1,ikr-ikz)
  e^{i\left(kz-\frac{\hbar k^2}{2m}t\right)}
\end{equation}
in terms of the confluent hypergeometric function ${}_1F_1$. Note
that for finite distances from the $z$-axis (i.e., for finite
coordinates $x$ and $y$), the asymptotic form of the Coulomb
continuum wavefunction for $z\rightarrow +\infty$ is just a plane
wave (without a logarithmic phase
term)~\cite{Landau,Merzbacher1998}.

Applying ACCTIVE to the outgoing-wave Coulomb problem, $t_r$ must be
chosen sufficiently long after $t$, so that each classical
trajectory $\tbr(t')$ propagates far enough towards the
$z\rightarrow+\infty$ asymptotic limit for the reference velocity to
become
\begin{equation}\label{eq:ICCWV}
  \bv_r \xrightarrow{~t_r\rightarrow\infty,~z\rightarrow+\infty~} \zu p/m ,
\end{equation}
in compliance with Eq.~(\ref{eq:CTInitial2}). In this and for the
following numerical example, we use as reference velocity the
initial trial velocity for points of the spatial numerical grid that
are sufficiently far away from the Coulomb singularity at the
origin.  The correct `current' velocities, $\bv\brt$ at the most
distant coordinates are subsequently used as trial velocities at the
nearest neighbor spatial grid points. This scheme is continued until
classical trajectories  for the entire spatiotemporal numerical grid
are calculated. Further details of the numerical calculation of
Coulomb wavefunctions within ACCTIVE are given in
Appendix~\ref{app:B}.

\begin{figure}
    \includegraphics[width=1.00\linewidth]{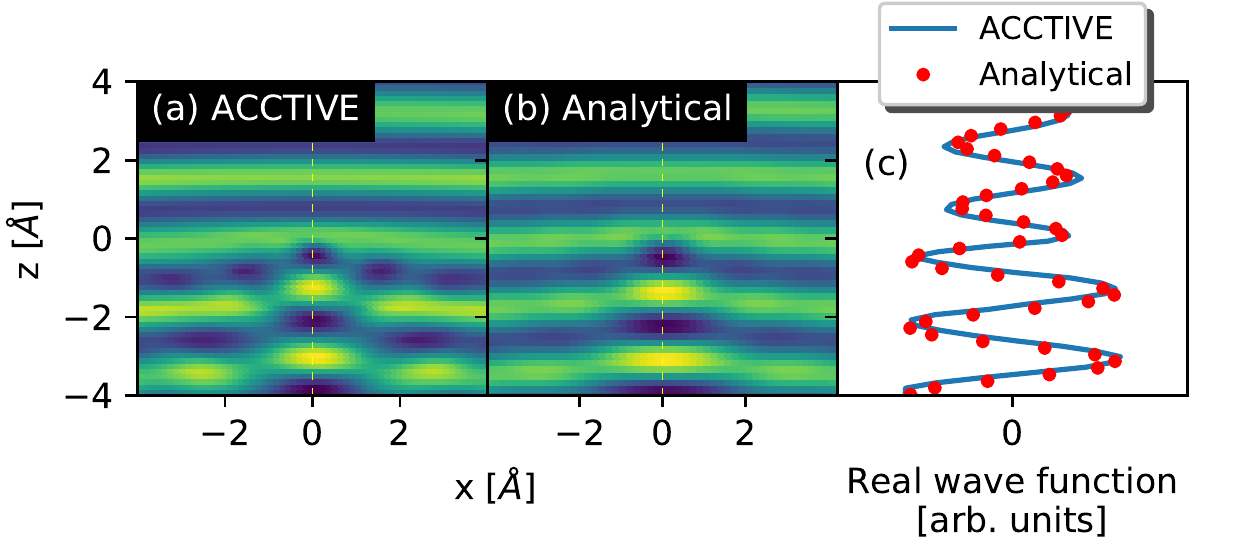}
    \caption{(Color online) Real part of an unbound Coulomb wavefunction ,
subject to the boundary condition given by an outgoing wave
propagating along the $z$-axis.
(a) Numerically calculated semi-classical 1st order ACCTIVE
wavefunction. (b) Analytical Coulomb wavefunction in the $y=0$
plane.
(c) Real part of the wavefunctions in (a) and (b) along the $z$-axis
for $x=y=0$.} \label{fig:2}
\end{figure}

Figure~\ref{fig:2} shows the very good agreement between the
numerically calculated 1st order ACCTIVE wavefunction~(\ref{eq:GV})
and the analytical Coulomb wavefunction (\ref{eq:CoulombAna}) for a
final electron kinetic energy of $p^2/2m=50$~eV. The color/gray
scale represents the real part of the wavefunction in the $x-z$
plane. Figures~\ref{fig:2}(a) and \ref{fig:2}(b) show the same
scattering pattern. Good quantitative agreement of the 1st order
ACCTIVE wavefunction and the analytical Coulomb wavefunction is
demonstrated in \rf{fig:2}(c).

\subsection{Coulomb-Volkov wavefunction} \label{sec:CV}

A more challenging third example is given by the motion of an
electron under the combined influence of a point charge (proton),
located at the coordinate origin, and a spatially homogeneous laser
pulse, subject to the boundary condition \re{eq:ICCW}. In this case,
the potentials in Eq.~(\ref{eq:TDSE}) are (in Coulomb
gauge~\cite{Merzbacher1998})
\begin{equation}\label{eq:potCV}
  \bA\brt = \bA(t) ,\quad \varphi\brt = -k_e\frac{e^2}{r}.
\end{equation}
Considering a laser pulse of finite duration, $t_r$ must be chosen
such that the laser electric field vanishes at $t_r$. This
combination of the two previous examples in Secs.~\ref{sec:Volkov}
and \ref{sec:Coulomb} constitutes the Coulomb-Volkov problem, for
which merely approximate
solutions~\cite{Rosenberg1993,Reiss1994CV,Macri2003CV,Uzer2019}, but
no analytical wavefunction are known. We assume a laser pulse with
15~eV central photon energy, a cosine-square temporal intensity
envelope with a pulse length of 0.5~fs full width at half intensity
maximum (FWHIM), and $3 \times 10^{15}$ W/cm$^2$ peak intensity. At
time $t=0$, the temporal pulse  profile is centered at $z=0$. We
enforce the outgoing-wave  boundary condition (\ref{eq:ICCW}) for an
asymptotic photoelectron kinetic energy of $p^2/2m=50$~eV. This
energy is reached at a sufficiently large distance of the outgoing
electron from the proton and long after the pulse has vanished.

In Fig.~\ref{fig:3} we compare the ACCTIVE-calculated Coulomb-Volkov
wavefunction with Coulomb and Volkov wavefunctions for identical
outgoing-wave boundary condition and 50~eV asymptotic photoelectron
kinetic energy. The Coulomb and Volkov wavefunctions are given for a
positive elementary charge and the same laser parameters as the
Coulomb-Volkov wave, respectively. The color/gray scale represents
the real part of the wavefunctions. We determined all numerical
parameters (numerical grid size, spacing and propagation time step)
to ensure convergence of the wavefunctions.

\begin{figure}
    \includegraphics[width=1.00\linewidth]{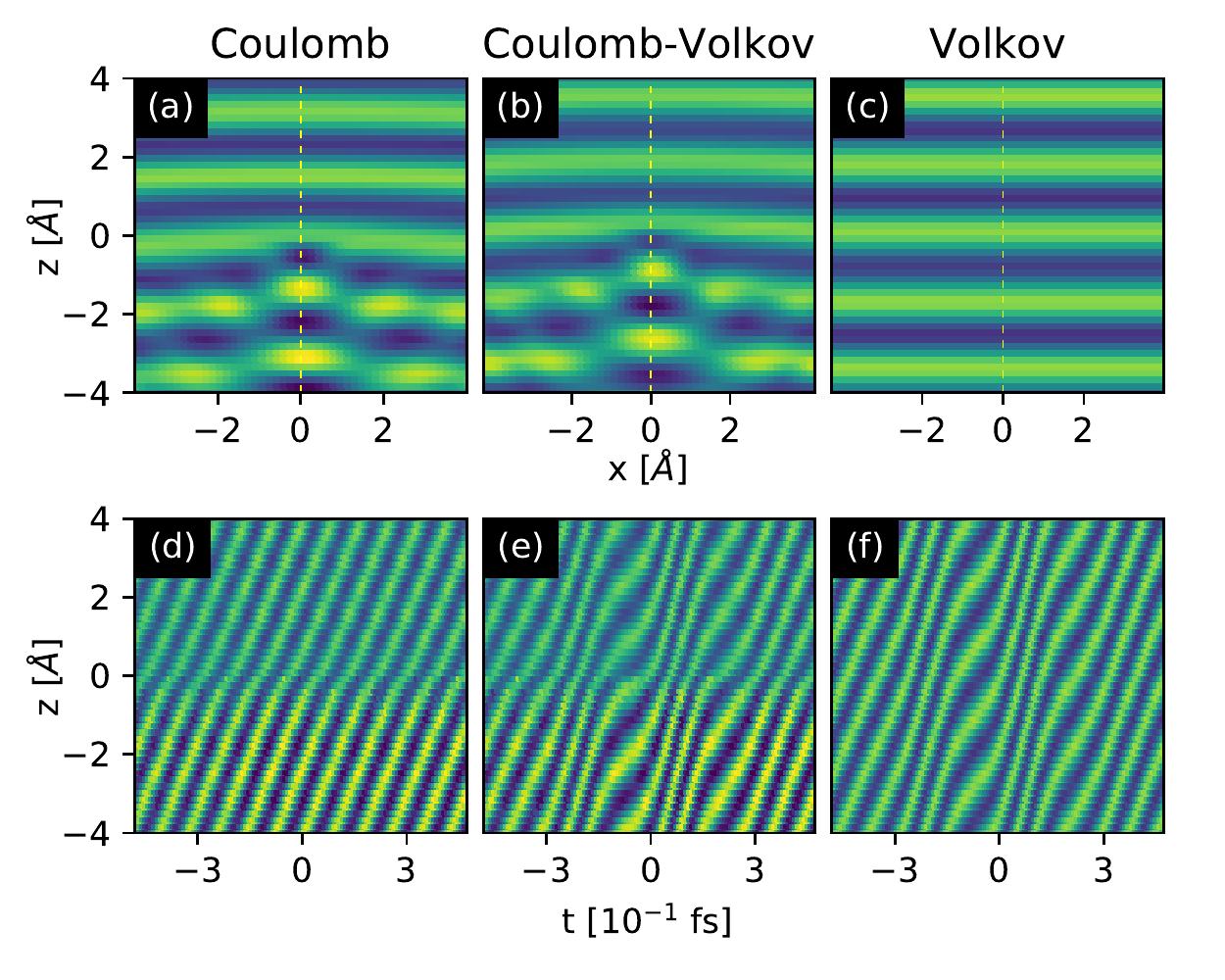}
\caption{(Color online) Real parts of (a,d) Coulomb, (b,e)
ACCTIVE-calculated Coulomb-Volkov, and (c,f) Volkov wavefunctions in
the $y=0$ plane. (a-c) Snapshots at time $t=0$, when the laser-pulse
center is at $z=0$. (d-f) Time evolution along the $z$-axis.}
\label{fig:3}
\end{figure}

Figures~\ref{fig:3}(a), \ref{fig:3}(b), and \ref{fig:3}(c), display
snapshots at time $t=0$ of the Coulomb, ACCTIVE-calculated
Coulomb-Volkov, and Volkov wavefunctions, respectively. The
Coulomb-Volkov wavefunction shows a similar (inverse) Coulomb
scattering pattern for the incident wave ($z<0$) as the Coulomb
wave. Its outgoing part ($z>0$) closely matches the phase of the
Volkov wave. On the other hand, the time-dependent evolution of the
Coulomb-Volkov wavefunction in the $y=0$ plane in
Fig.~\ref{fig:3}(e) shows laser-induced wavefront distortions -
similar to the Volkov wave in Fig.~\ref{fig:3}(f). The time
evolution of the ACCTIVE-calculated Coulomb-Volkov wavefunction
reveals the acceleration of the incoming and deceleration of the
outgoing wave near the proton at $z=0$ of the pure Coulomb wave in
Fig.~\ref{fig:3}(d). An animated version of this wavefunction
comparison can be found in the Supplemental
Material~\cite{SupplMat}.

\subsection{Streaked photoemission from hydrogen atoms} \label{sec:streak-atom}

As a fourth example, we employ ACCTIVE final-state wavefunctions to
calculate IR-streaked XUV photoelectron spectra from ground-state
hydrogen atoms~\cite{Thumm2015Chap}. We assume the ionizing XUV and
streaking IR pulse as linearly polarized along the $z$ axis. The
relative time delay between the centers of the two pulses, $\tau$,
is assumed positive in case the IR precedes XUV pulse. The electric
field $E_X(t)$ of the XUV pulse is characterized by a Gaussian
temporal profile, 55~eV central photon energy, and a pulse length of
200~as (FWHIM). The IR pulse has a cosine-squared temporal profile,
720 nm central wavelength, pulse duration of 2~fs FWHIM, and
$10^{11}$ W/cm$^2$ peak intensity.

We model streaked photoemission from the ground state of hydrogen,
$\left|\Psi_i\right>$, based on the quantum-mechanical transition
amplitude~\cite{Merzbacher1998,Liao2014,Thumm2015Chap,Li2018PRL}
\begin{equation} \label{eq:Tif}
 T(\bk_f,\tau) \sim \int dt~
   \big< \Psi_{\bk_f,\tau}^{C-V}
     \big|
       zE_X(t)
     \big|
     \Psi_i
   \big>,
\end{equation}
where the IR-pulse-dressed final state of the photoelectron,
$\big|\Psi_{\bk_f,\tau}^{C-V}\big>$, is a Coulomb-Volkov
wavefunction~\cite{Zhang2010} that we evaluate numerically using the
ACCTIVE method. In a comparison calculation, we replace the
Coulomb-Volkov state by the Volkov state $\big|\Psi^V_{\bk_f,\tau}
\big>$ and assume otherwise identical physical conditions. As
mentioned in the Introduction, the use of Volkov
states~\cite{Volkov1935} in photoionization calculations is referred
to as SFA~\cite{Thumm2015Chap} and amounts to neglecting the
interaction of the released photoelectron with the residual ion
(proton in the present case). We scrutinize streaked photoemission
spectra obtained with ACCTIVE-calculated Coulomb-Volkov final states
and in SFA against {\em ab initio} bench-mark calculations. In these
exact numerical calculations we directly solve the three-dimensional
TDSE using the SCID-TDSE time-propagation code~\cite{SCID-TDSE}.

Numerical results are shown in \rf{fig:4}. The streaked
photoemission spectra obtained with ACCTIVE-calculated
Coulomb-Volkov final states [Fig.~\ref{fig:4}(a)], in SFA
[Fig.~\ref{fig:4}(b)], and by direct numerical solution of the TDSE
[Fig.~\ref{fig:4}(c)] show very similar `streaking traces', i.e.,
oscillations of the asymptotic photoelectron energy with delay
$\tau$.
For a quantitative comparison, we plot in Fig.~\ref{fig:4}(d) the
centers of energy (CoEs) of the spectra in Figs.~\ref{fig:4}(a-c).
While the three calculations result in identical photoemission phase
shifts (streaking time delays) relative to the streaking IR field,
within the resolution of the graph, the ACCTIVE-calculated spectra
agree with the exact TDSE calculation, while the SFA calculation
predicts noticeably smaller CoEs due to the neglect of the Coulomb
potential in the final photoelectron state~\cite{Li2016}.

\begin{figure}
    \includegraphics[width=1.00\linewidth]{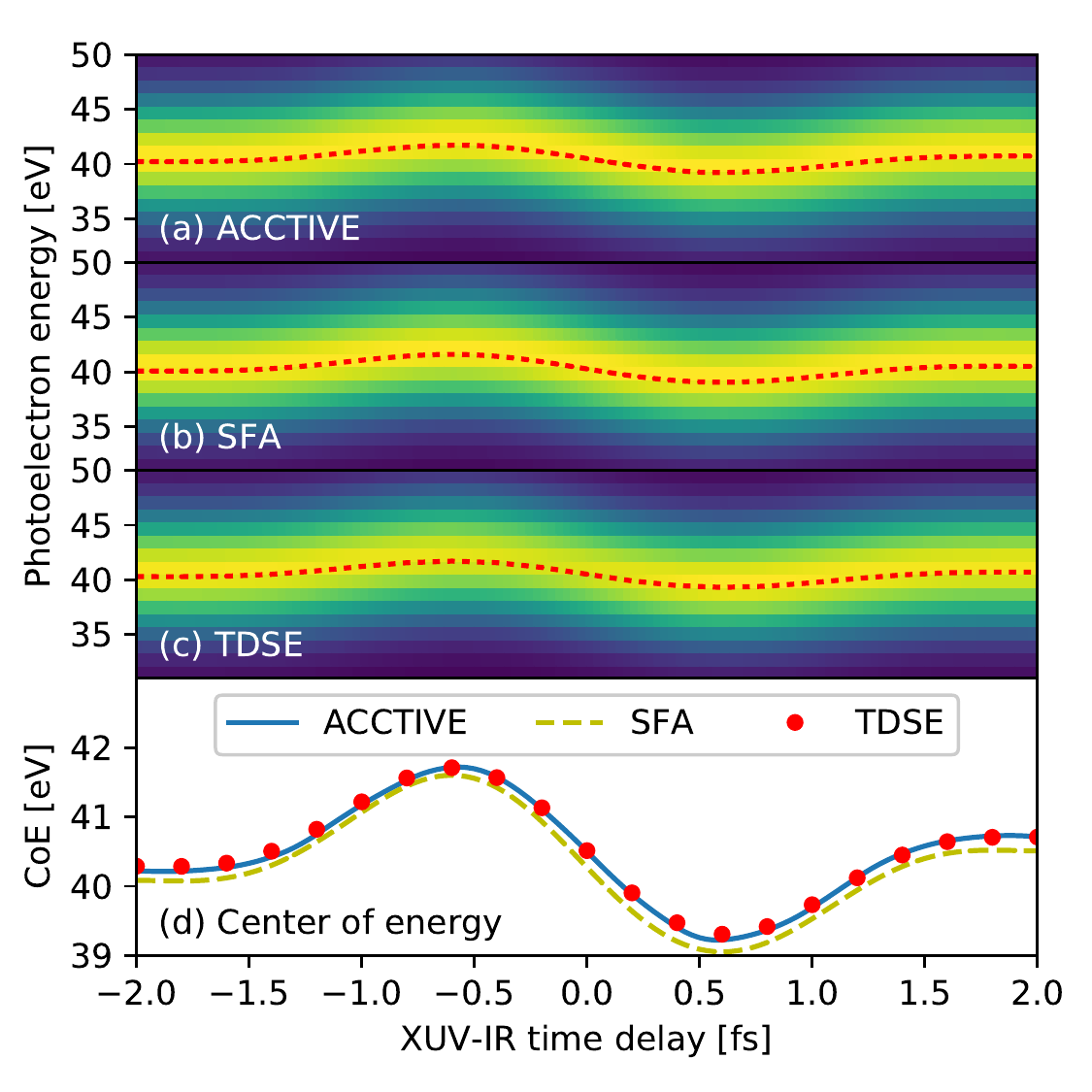}
\caption{(Color online) IR-streaked XUV photoelectron spectra, (a)
based on ACCTIVE-calculated Coulomb-Volkov final states, (b) in SFA,
and (c) obtained by direct numerical solution of the TDSE. Red
dotted lines in (a-c) indicate the respective centers of energy
(CoE). The spectral yields in (a-c) are normalized separately, to
their respective maxima.
(d) Comparison of the delay-dependent CoE for the spectra in (a-c).
} \label{fig:4}
\end{figure}

\subsection{Streaked photoemission from metal nanospheres} \label{sec:streak-np}

As a final, fifth, example, we apply the ACCTIVE method to model
photoelectron states in {\em spatially inhomogeneous}, plasmonically
enhanced IR electromagnetic fields. For this purpose, we investigate
streaked
photo\-emission~\cite{Li2016,Saydanzad2017,Li2017,Saydanzad2018} and
the reconstruction of plasmonic near-fields~\cite{Li2018PRL} for
gold nanospheres with a radius of $R=50$~nm. We represent the
electronic structure of the nanosphere in terms of eigenstates of a
square well with a potential depth of $V_0=-13.1$~eV and obtain the
photoelectron yield by incoherently adding the transition amplitudes
(\ref{eq:Tif}) over all occupied initial conduction-band
states~\cite{Zhang2009PRL,Liao2014,Thumm2015Chap}. For the
calculation of the transition amplitude (\ref{eq:Tif}) we closely
follow Ref.~\cite{Li2017}, with the important difference of
employing numerically calculated semiclassical ACCTIVE final
photoelectron wavefunctions, while in Ref.~\cite{Li2017} the SFA
approximation is used, applying heuristically generalized Volkov
final states and thus neglecting of the photoelectron interactions
with the residual nanoparticle.

For the ACCTIVE calculation we thus solve the TDSE~(\ref{eq:TDSE})
with the potentials
\begin{subequations}\label{eqs:potNP}
  \begin{eqnarray}
    \bA\brt &=& \int_t^\infty \bE_{\text{tot}}\brtp~dt' \\
    \varphi\brt &=&
      \begin{cases}
        V_0     &\quad r<R\\
        0       &\quad r \ge R
      \end{cases} \quad,
  \end{eqnarray}
\end{subequations}
and the boundary condition~\re{eq:ICCW}. Here, the asymptotic
wavefunction in \re{eq:ICCW} also serves as reference wavefunction
for the classical trajectory computation. The net time-dependent
inhomogeneous field $\bE_{\text{tot}}\brt$ is given by the
superposition of the homogeneous IR field of the incident streaking
pulse and the inhomogeneous plasmonic field produced by the
nanoparticle in response to the incident IR
pulse~\cite{Saydanzad2018,Li2018PRL}. For the streaking calculation,
we assume an XUV pulse with 30~eV central photon energy and Gaussian
temporal profile with a width of 200~as (FWHIM). We further suppose
a delayed Gaussian IR pulse with 720 nm central wavelength, 2.47 fs
(FWHIM) pulse length, and $5\times10^{10}$ W/cm$^2$ peak intensity.

\frf{fig:5} shows simulated streaked photoelectron spectra obtained
with ACCTIVE-calculated and Volkov final states for electron
emission along the XUV-pulse polarization direction. In this
direction, the effect of the induced plasmonic field on the
photoelectron is strongest~\cite{Li2018PRL}. The corresponding
spectra in Figs.~\ref{fig:5}(a) and \ref{fig:5}(b) show very similar
temporal oscillations of the photoelectron yield and CoE as a
function of both, asymptotic photoelectron energy and XUV-IR pulse
delay $\tau$. As for streaked photoemission from hydrogen atoms
discussed in Sec~\ref{sec:streak-atom} above, we find that the SFA
shifts the CoE to lower kinetic energies [Fig.~\ref{fig:5}(b), cf.
Fig.~\ref{fig:4}(d)]. Here, the SFA results in an approximately
1.5~eV lower CoE than the ACCTIVE calculation. This energy shift is
due to the fact that the SFA, by neglecting the potential well of
the nanosphere in the final photoelectron state, leads to an
unphysical enhancement of the photoemission cross section at lower
photoelectron kinetic energies, thereby increasing the weight of low
energy yields in the CoE average~\cite{Li2016}.

Addition comments on the comparison of streaked photoelectron
spectra within either ACCTIVE or based on Volkov wavefunctions can
be found in Appendix~\ref{app:C}.

\begin{figure}
    \includegraphics[width=1.00\linewidth]{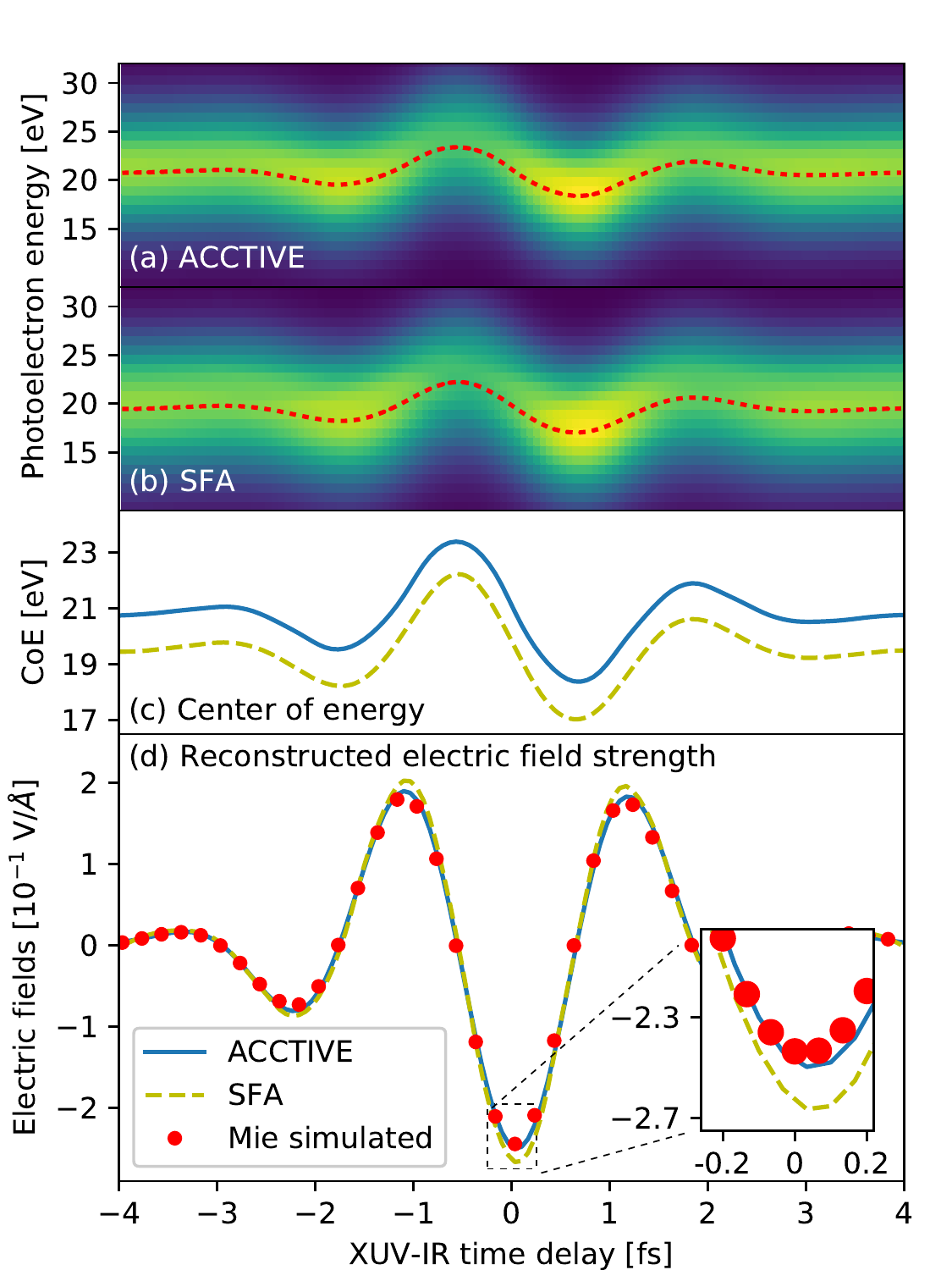}
\caption{(Color online) Simulated IR-streaked XUV photoelectron
spectra for photoemission along the XUV-pulse polarization direction
(a) using ACCTIVE final-states and (b) in SFA.
(c) Corresponding delay-dependent centers of energy.
(d) Comparison of the corresponding reconstructed plasmonic electric
near-fields at the point $(x,y,z) = (0,0,R)$ on the nanoparticle
surface with the Mie-theory-calculated electric field. }
\label{fig:5}
\end{figure}

From streaked photoemission spectra the plasmonic near-field at the
nanoparticle surface can be reconstructed as detailed in
Refs.~\cite{Li2018PRL,Saydanzad2018}. Figure~\ref{fig:5}(d) shows
the reconstructed net electric field $\bE_{\text{tot}}$ along the
XUV-pulse polarization direction, i.e., at the surface and on the
positive $z$ axis, of the nanosphere. The reconstruction of net
plasmonically enhanced near-fields from the simulated spectra in
Figs.~\ref{fig:5}(a) and \ref{fig:5}(b) was performed according to
the scheme proposed in Ref.~\cite{Li2018PRL}. The obtained
reconstructed fields are compared in Fig.~\ref{fig:5}(d) with the
net electric IR near-field obtained within Mie theory~\cite{Mie1908}
and used as input in the streaking calculations. As is seen in
Fig.~\ref{fig:5}(d), the ACCTIVE method improves the near-field
reconstruction in comparison with the SFA calculation. The
least-square deviation between the reconstructed and Mie-theory
calculated fields, assembled over the entire IR pulse length,
amounts to 1.62\% using the ACCTIVE wavefunction and 3.05\% using
the SFA. A comparative animation of reconstructed and analytical
electric fields at the surface of Au nanospheres can be found in the
Supplemental Material~\cite{SupplMat}.
The ACCTIVE method thus extends the applicability of the plasmonic
near-field reconstruction scheme in Ref.~\cite{Li2018PRL} to lower
XUV photon energies.

\section{Summary} \label{sec:conclusion}

In summary, we propose a semi-classical method, ACCTIVE, to solve
the TDSE for one active electron exposed to any spatially
inhomogeneous time-dependent external force field. We validate this
method by comparing ACCTIVE-calculated electronic wavefunctions with
known Coulomb and Volkov wavefunctions for the electronic dynamics
in Coulomb and intense laser fields, respectively, and by
scrutinizing ACCTIVE-calculated Coulomb-Volkov final photoelectron
wavefunctions (i) against {\em ab initio} numerical solutions of the
TDSE and (ii) in streaked photoemission from hydrogen atoms and
plasmonic metal nanospheres.

For streaked photoemission from hydrogen atoms, we demonstrate
excellent agreement of our ACCTIVE calculation with a benchmark {\em
ab initio} TDSE calculation, while a comparative calculation using
the SFA systematically deviates from the exact TDSE solution. For
streaked photoemission from Au nanospheres we find that ACCTIVE
final-state wavefunctions improve the reconstruction of plasmonic
near-fields over SFA calculations (based on Volkov final states) at
comparatively low photoelectron energies.

\appendix

\section{Derivation of \re{eq:Volkov}} \label{app:A}

We here derive  the Volkov wavefunction \re{eq:Volkov} using
ACCTIVE. Starting from the potentials and initial wavefunction in
\re{eq:ICV}, the velocity field along the classical trajectory
$\tilde{\br}(t)$ is
\begin{equation}
\bv\brt = \frac{\bp}{m} + \frac{q}{m}\int_{t_0}^t\bE(t') dt' =
\frac{\bp - q\bA(t)}{m}.
\end{equation}
Therefore,
\begin{equation}
  \tbr(t) = \br_0 + \int_{t_0}^t \bigg[\frac{\bp - q\bA(t)}{m}\bigg] dt',
\end{equation}
\begin{equation}
\nabla\cdot\bv(\br,t) = 0,
\end{equation}
and \re{eq:GV}, applied to the example in Sec.~\ref{sec:Volkov},
becomes
\begin{align}
 \Psi\brt
    &=\exp
      \Bigg\{
        \frac{i\bp\cdot\br_0}{\hbar} + \frac{i}{\hbar}\int_{t_0}^{t}
        \bigg[
          \frac{m}{2}
          \bigg(
            \frac{\bp - q\bA(t')}{m}
          \bigg)^2 \nonumber\\
          &\quad\quad + q
          \bigg(
            \frac{\bp - q\bA(t')}{m}
          \bigg) \cdot \bA(t')
        \bigg] dt'
      \Bigg\} \nonumber\\
    &=\exp
      \Bigg\{
        \frac{i\bp}{\hbar} \cdot
        \bigg[
          \br - \int_{t_0}^t \bigg(
            \frac{\bp - q\bA(t)}{m}
          \bigg) dt'
        \bigg] \nonumber\\
        &\quad\quad + \frac{i}{\hbar}\int_{t_0}^{t}
        \bigg[
          \frac{m}{2}
          \bigg(
            \frac{\bp - q\bA(t')}{m}
          \bigg)^2 \nonumber\\
          &\quad\quad + q
          \bigg(
            \frac{\bp - q\bA(t')}{m}
          \bigg) \cdot \bA(t')
        \bigg] dt'
      \Bigg\} \nonumber\\
    &=\exp
      \Bigg\{
        \frac{i\bp\cdot\br}{\hbar} + \frac{i}{\hbar}\int_{t_0}^{t}
        \bigg[
          \frac{m}{2}
          \bigg(
            \frac{\bp - q\bA(t')}{m}
          \bigg)^2 \nonumber\\
          &\quad\quad - m
          \bigg(
            \frac{\bp - q\bA(t')}{m}
          \bigg)^2
        \bigg] dt'
      \Bigg\} \nonumber\\
    &=\exp
        \bigg\{
          \frac{i~\bp\cdot\br}{\hbar}
          -\frac{i}{2m\hbar}\int_{t_0}^t
            \big[
              \bp - q\bA(t')
            \big]^2 dt'
        \bigg\}, \quad\quad
\end{align}
which is the Volkov wavefunction \re{eq:Volkov}.

\section{Numerical calculation of Coulomb wavefunctions using
ACCTIVE} \label{app:B}

The ACCTIVE method links a quantum-mechanical problem of obtaining
wavefunctions $\Psi\brt$ to a classical problem of determining
velocity fields $\bv\brt$. However, in some cases, e.g., for Coulomb
wavefunctions, such velocity fields are not uniquely defined
(\rf{fig:sup1}). This can result in interference patterns in the
obtained wavefunctions, as pointed out by Goldfarb \emph{et
al.}~\cite{Goldfarb2008}.

\begin{figure}
    \includegraphics[width=1.00\linewidth]{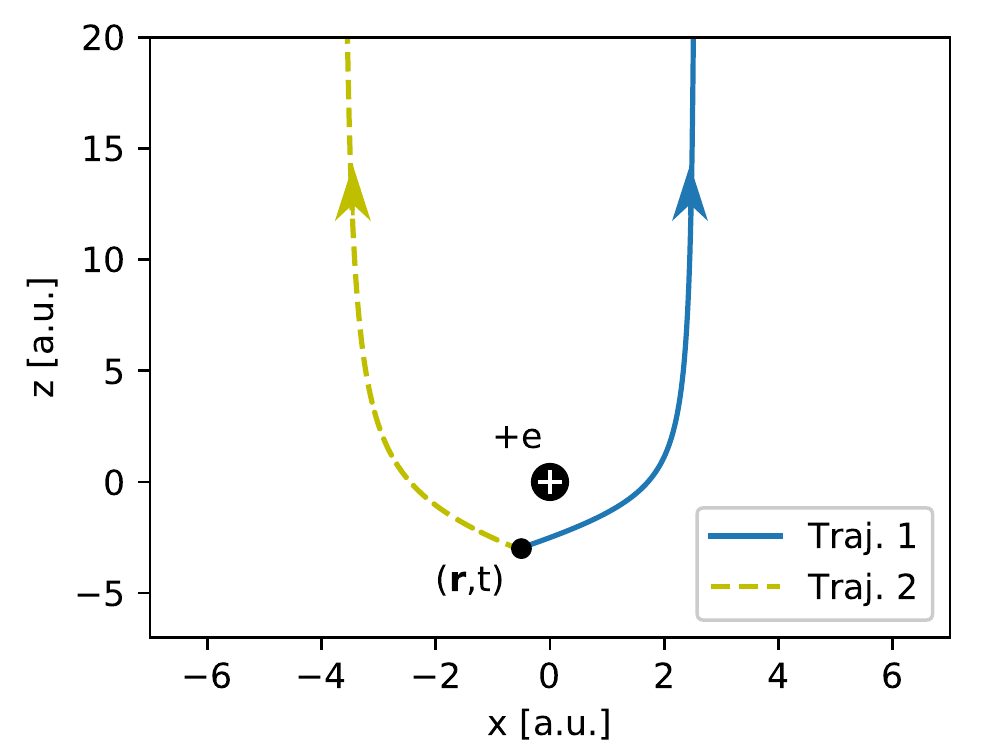}
    \caption{(Color online)
    Two possible classical trajectories passing through $\brt$ satisfying the
    same outgoing plane wave boundary condition. } \label{fig:sup1}
\end{figure}

For each event $\brt$, two possible classical trajectories can be
found to satisfy the same boundary condition of an outgoing plane
wave in \re{eq:ICCW}, as shown in \rf{fig:sup1}. Goldfarb \emph{et
al.}~\cite{Goldfarb2008} take this interference into account by
approximating the wavefunction as the superposition of contributions
from different trajectories,
\begin{equation}
  \Psi\brt \approx \sum_l \exp
    \Big[
      \frac{i}{\hbar}S^l
      \big(
        \tilde{\br}(t),t
      \big)
    \Big],
\end{equation}
where each action $S^l\brt$ is associated with a trajectory
$\tilde{\br}^l(t)$. In this work, we follow a different and simpler
approach.

The TDSE is a linear partial differential equation. Its solution can
be expressed as the superposition of a set of linearly independent
basis functions $\Psi^l\brt$,
\begin{equation}
  \Psi\brt = \sum_l C^l \Psi^l\brt = \sum_l C^l \exp
    \Big[
      \frac{i}{\hbar}S^l
      \big(
        \br,t
      \big)
    \Big],
\end{equation}
where each $S^l\brt$ is uniquely determined by a velocity field
$\bv^l\brt$ and the coefficients $C^l$ are obtained from the initial
condition,
\begin{equation}
  \Psi_0(\br) = \sum_l C^l \Psi^l(\br,t_0).
\end{equation}

\begin{figure}
    \includegraphics[width=1.00\linewidth]{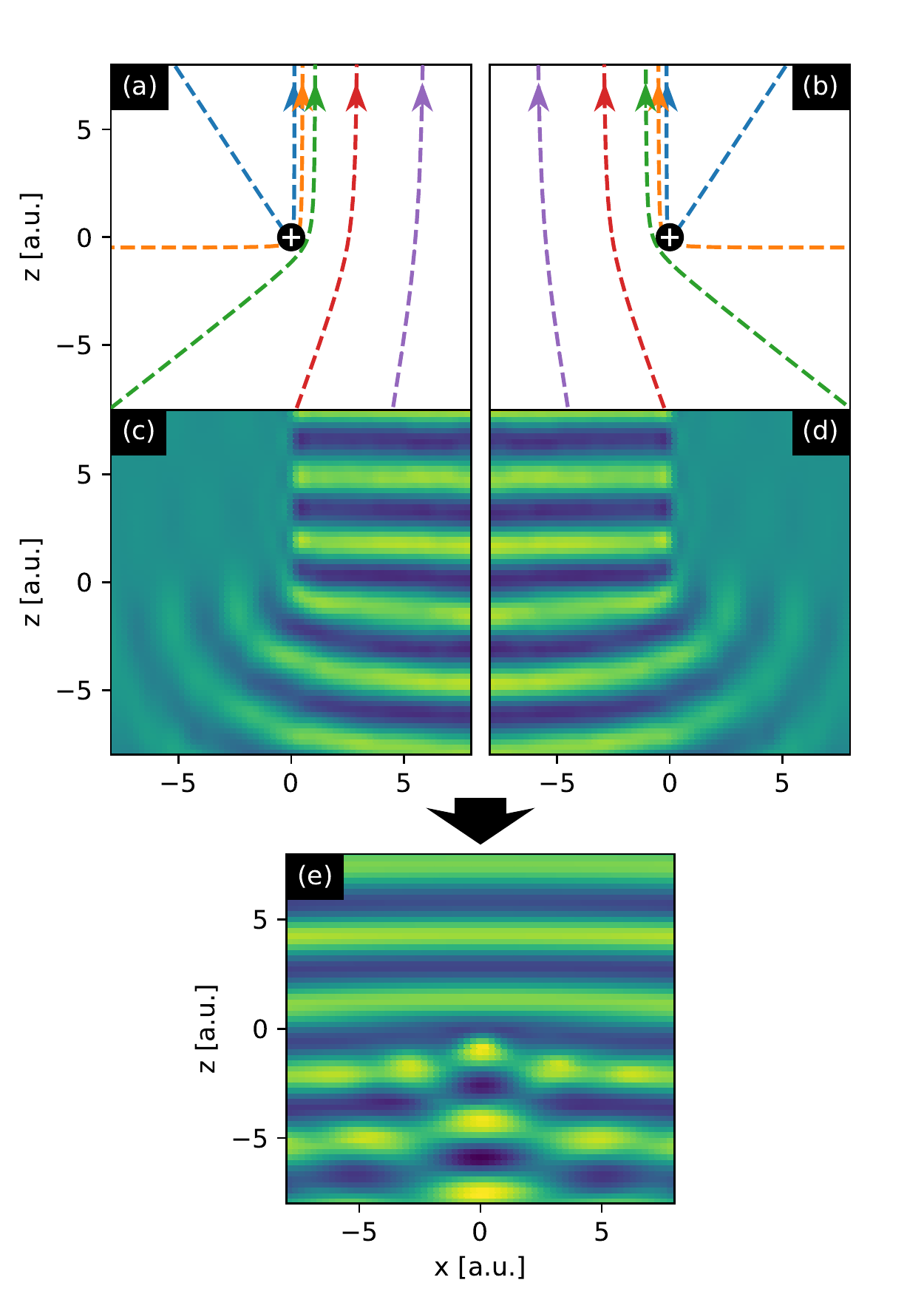}
    \caption{(Color online)
Two possible velocity fields (a) $\bv^+\brt$, and (b) $\bv^-\brt$.
(c) $\Psi^+\brt$ and (d) $\Psi^-\brt$ are the real parts of the
corresponding 1st order ACCTIVE wavefunctions at $y=0$ plane,
respectively, and (d) $\Psi\brt$ is the linear combination of these
two wavefunctions.} \label{fig:sup2}
\end{figure}

Since two possible trajectories can be obtained for each given event
$\brt$, we can find two  velocity fields, $\bv^+\brt$ and
$\bv^-\brt$, which are defined by
\begin{subequations} \label{eq:boundary}
  \begin{eqnarray}
    \bv^+(\br,t) &\xrightarrow{~z\rightarrow+\infty,~x>0~}& \zu p/m \\
    \bv^-(\br,t) &\xrightarrow{~z\rightarrow+\infty,~x<0~}& \zu p/m,
  \end{eqnarray}
\end{subequations}
as illustrated in \rf{fig:sup2}(a) and \ref{fig:sup2}(b),
respectively. Figures \ref{fig:sup2}(c) and \ref{fig:sup2}(d) show
the calculated 1st-order ACCTIVE wavefunctions, $\Psi^+\brt$ and
$\Psi^-\brt$, associated with these two velocity fields at $t=0$.
Numerical calculation shows that,
\begin{subequations}
  \begin{eqnarray}
    \Psi^+\brt &\xrightarrow{~z\rightarrow+\infty~}&
      \begin{cases}
        e^{ikz}     & x>0\\
        0           & x<0
      \end{cases}\\
    \Psi^-\brt &\xrightarrow{~z\rightarrow+\infty~}&
      \begin{cases}
        e^{ikz}     & x<0\\
        0           & x>0
      \end{cases}.
  \end{eqnarray}
\end{subequations}
Therefore, at $t_0$, $\Psi_0(\br) = \Psi(\br,0)$ can be written as
the linear combination of $\Psi^+(\br,t_0)$ and $\Psi^-(\br,t_0)$
and satisfies the boundary condition~(\ref{eq:boundary}),
\begin{equation}
  \Psi_0(\br) = \Psi^+(\br,t_0) + \Psi^-(\br,t_0).
\end{equation}
The wavefunction at any given time $t$ is then obtained with the
same coefficients,
\begin{equation}
  \Psi\brt = \Psi^+\brt + \Psi^-\brt,
\end{equation}
as shown in \rf{fig:sup2} (e).

\section{Comments on streaked photoemission from Au nanospheres} \label{app:C}

\frf{fig:5} in the main text shows the comparison of simulated
streaked photoelectron spectra using either ACCTIVE wavefunctions as
final states or Volkov wavefunction in SFA. ACCTIVE wavefunctions
are more accurate at low photoelectron energy, but entail higher
CoEs than Volkov wavefunctions [\rf{fig:5}(c)]. In comparison with
\rf{fig:4}(d), this might appear as counter-intuitive. An
explanation is given below.

\begin{figure}
    \includegraphics[width=1.00\linewidth]{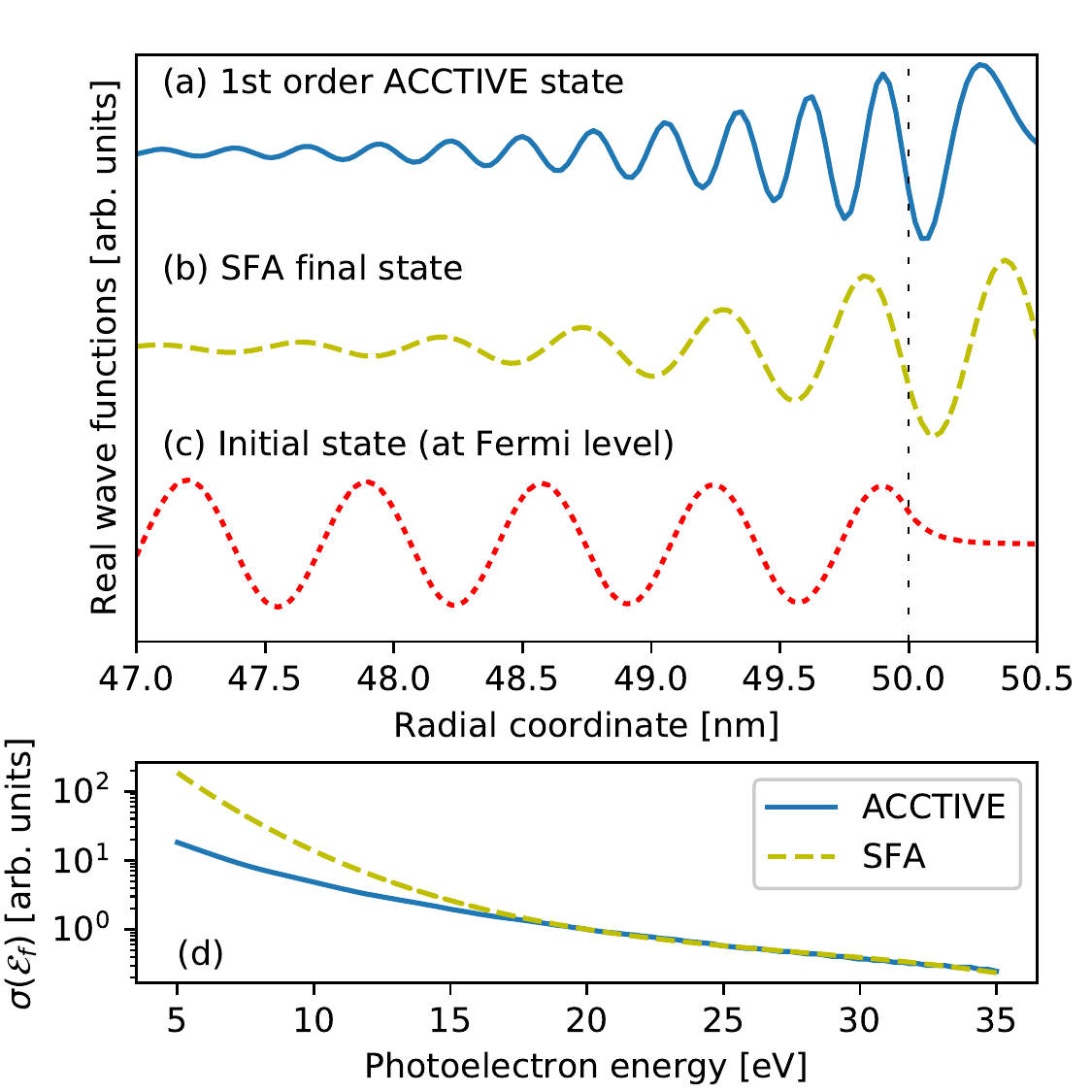}
    \caption{(Color online) Real parts of photoelectron final-state
wavefunctions near the surface of Au nanospheres along the XUV
polarization direction: (a) 1st order ACCTIVE wavefunction and (b)
SFA modeled wavefunction in Ref.~\cite{Li2018PRL}, for the electron
detection along the XUV polarization direction and asymptotic
photoelectron energy $\Ef=5$~eV. (c) Initial state wavefunction,
modeled as bound state in a spherical square well potential, at the
Fermi level. The vertical dashed line indicates the nanosphere
surface. (d) Simulated XUV photoemission cross sections. }
\label{fig:sup3}
\end{figure}

\frf{fig:sup3}(a) shows the real part of the 1st-order ACCTIVE
wavefunction near the Au nanosphere surface, and \rf{fig:sup3}(b)
the corresponding Volkov wavefunction in SFA~\cite{Li2018PRL}. Both
are calculated for photoelectron detection along the XUV
polarization direction and outgoing photoelectron energy $\Ef=5$~eV.
Inside the nanosphere, the Volkov final-state wavefunction neglects
the spherical well potential. It therefore has a longer wavelength
than the ACCTIVE wavefunction and more strongly overlaps with the
initial-state wavefunction shown in \rf{fig:sup3}(c). Thus, the
cross section, calculated following Ref.~\cite{Merzbacher1998}, is
larger in SFA than if based on ACCTIVE final states.

This effect becomes less significant a larger photoelectron kinetic
energies, where both, ACCTIVE and SFA wavefunctions have shorter
wavelengths and overlap less with initial-state wavefunction.
\frf{fig:sup3}(d) shows that the energy-dependent photoemission
cross sections calculated with ACCTIVE and Volkov final states
converge at large photoelectron energies, while at small energies
the SFA leads to larger cross sections. The net effect of this
cross-section difference is to put more weight on photoelectron
yields at lower energy and thus to shift streaking traces and CoEs
in SFA photoemission spectra to lower energies as compared to
ACCTIVE-calculated spectra.


\bigskip
\section*{Acknowledgement}

This work was supported in part by the Chemical Sciences,
Geosciences, and Biosciences Division, Office of Basic Energy
Sciences, Office of Science, US Department of Energy under Award
DEFG02-86ER13491 (attosecond interferometry), NSF grant no. PHY
1802085 (theory of photoemission from surfaces), and the Air Force
Office of Scientific Research award no. FA9550-17-1-0369
(recollision physics at the nanoscale).

\bibliography{GV}

\begin{thebibliography}{71}%
\makeatletter
\providecommand \@ifxundefined [1]{%
 \@ifx{#1\undefined}
}%
\providecommand \@ifnum [1]{%
 \ifnum #1\expandafter \@firstoftwo
 \else \expandafter \@secondoftwo
 \fi
}%
\providecommand \@ifx [1]{%
 \ifx #1\expandafter \@firstoftwo
 \else \expandafter \@secondoftwo
 \fi
}%
\providecommand \natexlab [1]{#1}%
\providecommand \enquote  [1]{``#1''}%
\providecommand \bibnamefont  [1]{#1}%
\providecommand \bibfnamefont [1]{#1}%
\providecommand \citenamefont [1]{#1}%
\providecommand \href@noop [0]{\@secondoftwo}%
\providecommand \href [0]{\begingroup \@sanitize@url \@href}%
\providecommand \@href[1]{\@@startlink{#1}\@@href}%
\providecommand \@@href[1]{\endgroup#1\@@endlink}%
\providecommand \@sanitize@url [0]{\catcode `\\12\catcode `\$12\catcode
  `\&12\catcode `\#12\catcode `\^12\catcode `\_12\catcode `\%12\relax}%
\providecommand \@@startlink[1]{}%
\providecommand \@@endlink[0]{}%
\providecommand \url  [0]{\begingroup\@sanitize@url \@url }%
\providecommand \@url [1]{\endgroup\@href {#1}{\urlprefix }}%
\providecommand \urlprefix  [0]{URL }%
\providecommand \Eprint [0]{\href }%
\providecommand \doibase [0]{http://dx.doi.org/}%
\providecommand \selectlanguage [0]{\@gobble}%
\providecommand \bibinfo  [0]{\@secondoftwo}%
\providecommand \bibfield  [0]{\@secondoftwo}%
\providecommand \translation [1]{[#1]}%
\providecommand \BibitemOpen [0]{}%
\providecommand \bibitemStop [0]{}%
\providecommand \bibitemNoStop [0]{.\EOS\space}%
\providecommand \EOS [0]{\spacefactor3000\relax}%
\providecommand \BibitemShut  [1]{\csname bibitem#1\endcsname}%
\let\auto@bib@innerbib\@empty
\bibitem [{\citenamefont {H\"ufner}(2003)}]{Hufner2003}%
  \BibitemOpen
  \bibfield  {author} {\bibinfo {author} {\bibfnamefont {S.}~\bibnamefont
  {H\"ufner}},\ }\href@noop {} {\emph {\bibinfo {title} {Photoelectron
  Spectroscopy. Principles and Applications}}}\ (\bibinfo  {publisher}
  {Springer},\ \bibinfo {address} {Berlin},\ \bibinfo {year}
  {2003})\BibitemShut {NoStop}%
\bibitem [{\citenamefont {Hentschel}\ \emph {et~al.}(2001)\citenamefont
  {Hentschel}, \citenamefont {Kienberger}, \citenamefont {Spielmann},
  \citenamefont {Reider}, \citenamefont {Milosevic}, \citenamefont {Brabec},
  \citenamefont {Corkum}, \citenamefont {Heinzmann}, \citenamefont {Drescher},\
  and\ \citenamefont {Krausz}}]{Hentschel2001}%
  \BibitemOpen
  \bibfield  {author} {\bibinfo {author} {\bibfnamefont {M.}~\bibnamefont
  {Hentschel}}, \bibinfo {author} {\bibfnamefont {R.}~\bibnamefont
  {Kienberger}}, \bibinfo {author} {\bibfnamefont {C.}~\bibnamefont
  {Spielmann}}, \bibinfo {author} {\bibfnamefont {G.~A.}\ \bibnamefont
  {Reider}}, \bibinfo {author} {\bibfnamefont {N.}~\bibnamefont {Milosevic}},
  \bibinfo {author} {\bibfnamefont {T.}~\bibnamefont {Brabec}}, \bibinfo
  {author} {\bibfnamefont {P.}~\bibnamefont {Corkum}}, \bibinfo {author}
  {\bibfnamefont {U.}~\bibnamefont {Heinzmann}}, \bibinfo {author}
  {\bibfnamefont {M.}~\bibnamefont {Drescher}}, \ and\ \bibinfo {author}
  {\bibfnamefont {F.}~\bibnamefont {Krausz}},\ }\href
  {https://doi.org/10.1038/35107000} {\bibfield  {journal} {\bibinfo  {journal}
  {Nature}\ }\textbf {\bibinfo {volume} {414}},\ \bibinfo {pages} {509}
  (\bibinfo {year} {2001})}\BibitemShut {NoStop}%
\bibitem [{\citenamefont {Chang}(2004)}]{Chang2004}%
  \BibitemOpen
  \bibfield  {author} {\bibinfo {author} {\bibfnamefont {Z.}~\bibnamefont
  {Chang}},\ }\href {\doibase 10.1103/PhysRevA.70.043802} {\bibfield  {journal}
  {\bibinfo  {journal} {Phys. Rev. A}\ }\textbf {\bibinfo {volume} {70}},\
  \bibinfo {pages} {043802} (\bibinfo {year} {2004})}\BibitemShut {NoStop}%
\bibitem [{\citenamefont {Sansone}\ \emph {et~al.}(2006)\citenamefont
  {Sansone}, \citenamefont {Benedetti}, \citenamefont {Calegari}, \citenamefont
  {Vozzi}, \citenamefont {Avaldi}, \citenamefont {Flammini}, \citenamefont
  {Poletto}, \citenamefont {Villoresi}, \citenamefont {Altucci}, \citenamefont
  {Velotta}, \citenamefont {Stagira}, \citenamefont {De~Silvestri},\ and\
  \citenamefont {Nisoli}}]{Sansone2006}%
  \BibitemOpen
  \bibfield  {author} {\bibinfo {author} {\bibfnamefont {G.}~\bibnamefont
  {Sansone}}, \bibinfo {author} {\bibfnamefont {E.}~\bibnamefont {Benedetti}},
  \bibinfo {author} {\bibfnamefont {F.}~\bibnamefont {Calegari}}, \bibinfo
  {author} {\bibfnamefont {C.}~\bibnamefont {Vozzi}}, \bibinfo {author}
  {\bibfnamefont {L.}~\bibnamefont {Avaldi}}, \bibinfo {author} {\bibfnamefont
  {R.}~\bibnamefont {Flammini}}, \bibinfo {author} {\bibfnamefont
  {L.}~\bibnamefont {Poletto}}, \bibinfo {author} {\bibfnamefont
  {P.}~\bibnamefont {Villoresi}}, \bibinfo {author} {\bibfnamefont
  {C.}~\bibnamefont {Altucci}}, \bibinfo {author} {\bibfnamefont
  {R.}~\bibnamefont {Velotta}}, \bibinfo {author} {\bibfnamefont
  {S.}~\bibnamefont {Stagira}}, \bibinfo {author} {\bibfnamefont
  {S.}~\bibnamefont {De~Silvestri}}, \ and\ \bibinfo {author} {\bibfnamefont
  {M.}~\bibnamefont {Nisoli}},\ }\href {\doibase 10.1126/science.1132838}
  {\bibfield  {journal} {\bibinfo  {journal} {Science}\ }\textbf {\bibinfo
  {volume} {314}},\ \bibinfo {pages} {443} (\bibinfo {year}
  {2006})}\BibitemShut {NoStop}%
\bibitem [{\citenamefont {Krausz}\ and\ \citenamefont
  {Ivanov}(2009)}]{Krausz2009}%
  \BibitemOpen
  \bibfield  {author} {\bibinfo {author} {\bibfnamefont {F.}~\bibnamefont
  {Krausz}}\ and\ \bibinfo {author} {\bibfnamefont {M.}~\bibnamefont
  {Ivanov}},\ }\href {\doibase 10.1103/RevModPhys.81.163} {\bibfield  {journal}
  {\bibinfo  {journal} {Rev. Mod. Phys.}\ }\textbf {\bibinfo {volume} {81}},\
  \bibinfo {pages} {163} (\bibinfo {year} {2009})}\BibitemShut {NoStop}%
\bibitem [{\citenamefont {Thumm}\ \emph {et~al.}(2015)\citenamefont {Thumm},
  \citenamefont {Liao}, \citenamefont {Bothschafter}, \citenamefont
  {S\"u\ss{}mann}, \citenamefont {Kling},\ and\ \citenamefont
  {Kienberger}}]{Thumm2015Chap}%
  \BibitemOpen
  \bibfield  {author} {\bibinfo {author} {\bibfnamefont {U.}~\bibnamefont
  {Thumm}}, \bibinfo {author} {\bibfnamefont {Q.}~\bibnamefont {Liao}},
  \bibinfo {author} {\bibfnamefont {E.~M.}\ \bibnamefont {Bothschafter}},
  \bibinfo {author} {\bibfnamefont {F.}~\bibnamefont {S\"u\ss{}mann}}, \bibinfo
  {author} {\bibfnamefont {M.~F.}\ \bibnamefont {Kling}}, \ and\ \bibinfo
  {author} {\bibfnamefont {R.}~\bibnamefont {Kienberger}},\ }in\ \href@noop {}
  {\emph {\bibinfo {booktitle} {The Oxford Handbook of Innovation}}},\ \bibinfo
  {editor} {edited by\ \bibinfo {editor} {\bibfnamefont {D.}~\bibnamefont
  {Andrew}}}\ (\bibinfo  {publisher} {Wiley},\ \bibinfo {address} {New York},\
  \bibinfo {year} {2015})\ Chap.~\bibinfo {chapter} {13}\BibitemShut {NoStop}%
\bibitem [{\citenamefont {Paul}\ \emph {et~al.}(2001)\citenamefont {Paul},
  \citenamefont {Toma}, \citenamefont {Breger}, \citenamefont {Mullot},
  \citenamefont {Aug{\'e}}, \citenamefont {Balcou}, \citenamefont {Muller},\
  and\ \citenamefont {Agostini}}]{Paul2001}%
  \BibitemOpen
  \bibfield  {author} {\bibinfo {author} {\bibfnamefont {P.~M.}\ \bibnamefont
  {Paul}}, \bibinfo {author} {\bibfnamefont {E.~S.}\ \bibnamefont {Toma}},
  \bibinfo {author} {\bibfnamefont {P.}~\bibnamefont {Breger}}, \bibinfo
  {author} {\bibfnamefont {G.}~\bibnamefont {Mullot}}, \bibinfo {author}
  {\bibfnamefont {F.}~\bibnamefont {Aug{\'e}}}, \bibinfo {author}
  {\bibfnamefont {P.}~\bibnamefont {Balcou}}, \bibinfo {author} {\bibfnamefont
  {H.~G.}\ \bibnamefont {Muller}}, \ and\ \bibinfo {author} {\bibfnamefont
  {P.}~\bibnamefont {Agostini}},\ }\href {\doibase 10.1126/science.1059413}
  {\bibfield  {journal} {\bibinfo  {journal} {Science}\ }\textbf {\bibinfo
  {volume} {292}},\ \bibinfo {pages} {1689} (\bibinfo {year}
  {2001})}\BibitemShut {NoStop}%
\bibitem [{\citenamefont {Kl\"under}\ \emph {et~al.}(2011)\citenamefont
  {Kl\"under}, \citenamefont {Dahlstr\"om}, \citenamefont {Gisselbrecht},
  \citenamefont {Fordell}, \citenamefont {Swoboda}, \citenamefont {Gu\'enot},
  \citenamefont {Johnsson}, \citenamefont {Caillat}, \citenamefont
  {Mauritsson}, \citenamefont {Maquet}, \citenamefont {Ta\"{\i}eb},\ and\
  \citenamefont {L'Huillier}}]{Klunder2011}%
  \BibitemOpen
  \bibfield  {author} {\bibinfo {author} {\bibfnamefont {K.}~\bibnamefont
  {Kl\"under}}, \bibinfo {author} {\bibfnamefont {J.~M.}\ \bibnamefont
  {Dahlstr\"om}}, \bibinfo {author} {\bibfnamefont {M.}~\bibnamefont
  {Gisselbrecht}}, \bibinfo {author} {\bibfnamefont {T.}~\bibnamefont
  {Fordell}}, \bibinfo {author} {\bibfnamefont {M.}~\bibnamefont {Swoboda}},
  \bibinfo {author} {\bibfnamefont {D.}~\bibnamefont {Gu\'enot}}, \bibinfo
  {author} {\bibfnamefont {P.}~\bibnamefont {Johnsson}}, \bibinfo {author}
  {\bibfnamefont {J.}~\bibnamefont {Caillat}}, \bibinfo {author} {\bibfnamefont
  {J.}~\bibnamefont {Mauritsson}}, \bibinfo {author} {\bibfnamefont
  {A.}~\bibnamefont {Maquet}}, \bibinfo {author} {\bibfnamefont
  {R.}~\bibnamefont {Ta\"{\i}eb}}, \ and\ \bibinfo {author} {\bibfnamefont
  {A.}~\bibnamefont {L'Huillier}},\ }\href {\doibase
  10.1103/PhysRevLett.106.143002} {\bibfield  {journal} {\bibinfo  {journal}
  {Phys. Rev. Lett.}\ }\textbf {\bibinfo {volume} {106}},\ \bibinfo {pages}
  {143002} (\bibinfo {year} {2011})}\BibitemShut {NoStop}%
\bibitem [{\citenamefont {Locher}\ \emph {et~al.}(2015)\citenamefont {Locher},
  \citenamefont {Castiglioni}, \citenamefont {Lucchini}, \citenamefont {Greif},
  \citenamefont {Gallmann}, \citenamefont {Osterwalder}, \citenamefont
  {Hengsberger},\ and\ \citenamefont {Keller}}]{Locher2015}%
  \BibitemOpen
  \bibfield  {author} {\bibinfo {author} {\bibfnamefont {R.}~\bibnamefont
  {Locher}}, \bibinfo {author} {\bibfnamefont {L.}~\bibnamefont {Castiglioni}},
  \bibinfo {author} {\bibfnamefont {M.}~\bibnamefont {Lucchini}}, \bibinfo
  {author} {\bibfnamefont {M.}~\bibnamefont {Greif}}, \bibinfo {author}
  {\bibfnamefont {L.}~\bibnamefont {Gallmann}}, \bibinfo {author}
  {\bibfnamefont {J.}~\bibnamefont {Osterwalder}}, \bibinfo {author}
  {\bibfnamefont {M.}~\bibnamefont {Hengsberger}}, \ and\ \bibinfo {author}
  {\bibfnamefont {U.}~\bibnamefont {Keller}},\ }\href {\doibase
  10.1364/OPTICA.2.000405} {\bibfield  {journal} {\bibinfo  {journal} {Optica}\
  }\textbf {\bibinfo {volume} {2}},\ \bibinfo {pages} {405} (\bibinfo {year}
  {2015})}\BibitemShut {NoStop}%
\bibitem [{\citenamefont {Drescher}\ \emph {et~al.}(2002)\citenamefont
  {Drescher}, \citenamefont {Hentschel}, \citenamefont {Kienberger},
  \citenamefont {Uiberacker}, \citenamefont {Yakovlev}, \citenamefont
  {Scrinzi}, \citenamefont {Westerwalbesloh}, \citenamefont {Kleineberg},
  \citenamefont {Heinzmann},\ and\ \citenamefont {Krausz}}]{Drescher2002}%
  \BibitemOpen
  \bibfield  {author} {\bibinfo {author} {\bibfnamefont {M.}~\bibnamefont
  {Drescher}}, \bibinfo {author} {\bibfnamefont {M.}~\bibnamefont {Hentschel}},
  \bibinfo {author} {\bibfnamefont {R.}~\bibnamefont {Kienberger}}, \bibinfo
  {author} {\bibfnamefont {M.}~\bibnamefont {Uiberacker}}, \bibinfo {author}
  {\bibfnamefont {V.}~\bibnamefont {Yakovlev}}, \bibinfo {author}
  {\bibfnamefont {A.}~\bibnamefont {Scrinzi}}, \bibinfo {author} {\bibfnamefont
  {T.}~\bibnamefont {Westerwalbesloh}}, \bibinfo {author} {\bibfnamefont
  {U.}~\bibnamefont {Kleineberg}}, \bibinfo {author} {\bibfnamefont
  {U.}~\bibnamefont {Heinzmann}}, \ and\ \bibinfo {author} {\bibfnamefont
  {F.}~\bibnamefont {Krausz}},\ }\href {https://doi.org/10.1038/nature01143}
  {\bibfield  {journal} {\bibinfo  {journal} {Nature}\ }\textbf {\bibinfo
  {volume} {419}},\ \bibinfo {pages} {803} (\bibinfo {year}
  {2002})}\BibitemShut {NoStop}%
\bibitem [{\citenamefont {Kienberger}\ \emph {et~al.}(2004)\citenamefont
  {Kienberger}, \citenamefont {Goulielmakis}, \citenamefont {Uiberacker},
  \citenamefont {Baltuska}, \citenamefont {Yakovlev}, \citenamefont {Bammer},
  \citenamefont {Scrinzi}, \citenamefont {Westerwalbesloh}, \citenamefont
  {Kleineberg}, \citenamefont {Heinzmann}, \citenamefont {Drescher},\ and\
  \citenamefont {Krausz}}]{Kienberger2004}%
  \BibitemOpen
  \bibfield  {author} {\bibinfo {author} {\bibfnamefont {R.}~\bibnamefont
  {Kienberger}}, \bibinfo {author} {\bibfnamefont {E.}~\bibnamefont
  {Goulielmakis}}, \bibinfo {author} {\bibfnamefont {M.}~\bibnamefont
  {Uiberacker}}, \bibinfo {author} {\bibfnamefont {A.}~\bibnamefont
  {Baltuska}}, \bibinfo {author} {\bibfnamefont {V.}~\bibnamefont {Yakovlev}},
  \bibinfo {author} {\bibfnamefont {F.}~\bibnamefont {Bammer}}, \bibinfo
  {author} {\bibfnamefont {A.}~\bibnamefont {Scrinzi}}, \bibinfo {author}
  {\bibfnamefont {T.}~\bibnamefont {Westerwalbesloh}}, \bibinfo {author}
  {\bibfnamefont {U.}~\bibnamefont {Kleineberg}}, \bibinfo {author}
  {\bibfnamefont {U.}~\bibnamefont {Heinzmann}}, \bibinfo {author}
  {\bibfnamefont {M.}~\bibnamefont {Drescher}}, \ and\ \bibinfo {author}
  {\bibfnamefont {F.}~\bibnamefont {Krausz}},\ }\href
  {https://doi.org/10.1038/nature02277} {\bibfield  {journal} {\bibinfo
  {journal} {Nature}\ }\textbf {\bibinfo {volume} {427}},\ \bibinfo {pages}
  {817} (\bibinfo {year} {2004})}\BibitemShut {NoStop}%
\bibitem [{\citenamefont {Johnsson}\ \emph {et~al.}(2007)\citenamefont
  {Johnsson}, \citenamefont {Mauritsson}, \citenamefont {Remetter},
  \citenamefont {L'Huillier},\ and\ \citenamefont {Schafer}}]{Johnsson2007}%
  \BibitemOpen
  \bibfield  {author} {\bibinfo {author} {\bibfnamefont {P.}~\bibnamefont
  {Johnsson}}, \bibinfo {author} {\bibfnamefont {J.}~\bibnamefont
  {Mauritsson}}, \bibinfo {author} {\bibfnamefont {T.}~\bibnamefont
  {Remetter}}, \bibinfo {author} {\bibfnamefont {A.}~\bibnamefont
  {L'Huillier}}, \ and\ \bibinfo {author} {\bibfnamefont {K.~J.}\ \bibnamefont
  {Schafer}},\ }\href {\doibase 10.1103/PhysRevLett.99.233001} {\bibfield
  {journal} {\bibinfo  {journal} {Phys. Rev. Lett.}\ }\textbf {\bibinfo
  {volume} {99}},\ \bibinfo {pages} {233001} (\bibinfo {year}
  {2007})}\BibitemShut {NoStop}%
\bibitem [{\citenamefont {Wang}\ \emph {et~al.}(2010)\citenamefont {Wang},
  \citenamefont {Chini}, \citenamefont {Chen}, \citenamefont {Zhang},
  \citenamefont {He}, \citenamefont {Cheng}, \citenamefont {Wu}, \citenamefont
  {Thumm},\ and\ \citenamefont {Chang}}]{Wang2010}%
  \BibitemOpen
  \bibfield  {author} {\bibinfo {author} {\bibfnamefont {H.}~\bibnamefont
  {Wang}}, \bibinfo {author} {\bibfnamefont {M.}~\bibnamefont {Chini}},
  \bibinfo {author} {\bibfnamefont {S.}~\bibnamefont {Chen}}, \bibinfo {author}
  {\bibfnamefont {C.-H.}\ \bibnamefont {Zhang}}, \bibinfo {author}
  {\bibfnamefont {F.}~\bibnamefont {He}}, \bibinfo {author} {\bibfnamefont
  {Y.}~\bibnamefont {Cheng}}, \bibinfo {author} {\bibfnamefont
  {Y.}~\bibnamefont {Wu}}, \bibinfo {author} {\bibfnamefont {U.}~\bibnamefont
  {Thumm}}, \ and\ \bibinfo {author} {\bibfnamefont {Z.}~\bibnamefont
  {Chang}},\ }\href {\doibase 10.1103/PhysRevLett.105.143002} {\bibfield
  {journal} {\bibinfo  {journal} {Phys. Rev. Lett.}\ }\textbf {\bibinfo
  {volume} {105}},\ \bibinfo {pages} {143002} (\bibinfo {year}
  {2010})}\BibitemShut {NoStop}%
\bibitem [{\citenamefont {Schultze}\ \emph {et~al.}(2010)\citenamefont
  {Schultze}, \citenamefont {Fie{\ss}}, \citenamefont {Karpowicz},
  \citenamefont {Gagnon}, \citenamefont {Korbman}, \citenamefont {Hofstetter},
  \citenamefont {Neppl}, \citenamefont {Cavalieri}, \citenamefont {Komninos},
  \citenamefont {Mercouris}, \citenamefont {Nicolaides}, \citenamefont
  {Pazourek}, \citenamefont {Nagele}, \citenamefont {Feist}, \citenamefont
  {Burgd{\"o}rfer}, \citenamefont {Azzeer}, \citenamefont {Ernstorfer},
  \citenamefont {Kienberger}, \citenamefont {Kleineberg}, \citenamefont
  {Goulielmakis}, \citenamefont {Krausz},\ and\ \citenamefont
  {Yakovlev}}]{Schultze2010}%
  \BibitemOpen
  \bibfield  {author} {\bibinfo {author} {\bibfnamefont {M.}~\bibnamefont
  {Schultze}}, \bibinfo {author} {\bibfnamefont {M.}~\bibnamefont {Fie{\ss}}},
  \bibinfo {author} {\bibfnamefont {N.}~\bibnamefont {Karpowicz}}, \bibinfo
  {author} {\bibfnamefont {J.}~\bibnamefont {Gagnon}}, \bibinfo {author}
  {\bibfnamefont {M.}~\bibnamefont {Korbman}}, \bibinfo {author} {\bibfnamefont
  {M.}~\bibnamefont {Hofstetter}}, \bibinfo {author} {\bibfnamefont
  {S.}~\bibnamefont {Neppl}}, \bibinfo {author} {\bibfnamefont {A.~L.}\
  \bibnamefont {Cavalieri}}, \bibinfo {author} {\bibfnamefont {Y.}~\bibnamefont
  {Komninos}}, \bibinfo {author} {\bibfnamefont {T.}~\bibnamefont {Mercouris}},
  \bibinfo {author} {\bibfnamefont {C.~A.}\ \bibnamefont {Nicolaides}},
  \bibinfo {author} {\bibfnamefont {R.}~\bibnamefont {Pazourek}}, \bibinfo
  {author} {\bibfnamefont {S.}~\bibnamefont {Nagele}}, \bibinfo {author}
  {\bibfnamefont {J.}~\bibnamefont {Feist}}, \bibinfo {author} {\bibfnamefont
  {J.}~\bibnamefont {Burgd{\"o}rfer}}, \bibinfo {author} {\bibfnamefont
  {A.~M.}\ \bibnamefont {Azzeer}}, \bibinfo {author} {\bibfnamefont
  {R.}~\bibnamefont {Ernstorfer}}, \bibinfo {author} {\bibfnamefont
  {R.}~\bibnamefont {Kienberger}}, \bibinfo {author} {\bibfnamefont
  {U.}~\bibnamefont {Kleineberg}}, \bibinfo {author} {\bibfnamefont
  {E.}~\bibnamefont {Goulielmakis}}, \bibinfo {author} {\bibfnamefont
  {F.}~\bibnamefont {Krausz}}, \ and\ \bibinfo {author} {\bibfnamefont {V.~S.}\
  \bibnamefont {Yakovlev}},\ }\href {\doibase 10.1126/science.1189401}
  {\bibfield  {journal} {\bibinfo  {journal} {Science}\ }\textbf {\bibinfo
  {volume} {328}},\ \bibinfo {pages} {1658} (\bibinfo {year}
  {2010})}\BibitemShut {NoStop}%
\bibitem [{\citenamefont {Ott}\ \emph {et~al.}(2013)\citenamefont {Ott},
  \citenamefont {Kaldun}, \citenamefont {Raith}, \citenamefont {Meyer},
  \citenamefont {Laux}, \citenamefont {Evers}, \citenamefont {Keitel},
  \citenamefont {Greene},\ and\ \citenamefont {Pfeifer}}]{Ott2013}%
  \BibitemOpen
  \bibfield  {author} {\bibinfo {author} {\bibfnamefont {C.}~\bibnamefont
  {Ott}}, \bibinfo {author} {\bibfnamefont {A.}~\bibnamefont {Kaldun}},
  \bibinfo {author} {\bibfnamefont {P.}~\bibnamefont {Raith}}, \bibinfo
  {author} {\bibfnamefont {K.}~\bibnamefont {Meyer}}, \bibinfo {author}
  {\bibfnamefont {M.}~\bibnamefont {Laux}}, \bibinfo {author} {\bibfnamefont
  {J.}~\bibnamefont {Evers}}, \bibinfo {author} {\bibfnamefont {C.~H.}\
  \bibnamefont {Keitel}}, \bibinfo {author} {\bibfnamefont {C.~H.}\
  \bibnamefont {Greene}}, \ and\ \bibinfo {author} {\bibfnamefont
  {T.}~\bibnamefont {Pfeifer}},\ }\href {\doibase 10.1126/science.1234407}
  {\bibfield  {journal} {\bibinfo  {journal} {Science}\ }\textbf {\bibinfo
  {volume} {340}},\ \bibinfo {pages} {716} (\bibinfo {year}
  {2013})}\BibitemShut {NoStop}%
\bibitem [{\citenamefont {Bernhardt}\ \emph {et~al.}(2014)\citenamefont
  {Bernhardt}, \citenamefont {Beck}, \citenamefont {Li}, \citenamefont
  {Warrick}, \citenamefont {Bell}, \citenamefont {Haxton}, \citenamefont
  {McCurdy}, \citenamefont {Neumark},\ and\ \citenamefont
  {Leone}}]{Bernhard2014}%
  \BibitemOpen
  \bibfield  {author} {\bibinfo {author} {\bibfnamefont {B.}~\bibnamefont
  {Bernhardt}}, \bibinfo {author} {\bibfnamefont {A.~R.}\ \bibnamefont {Beck}},
  \bibinfo {author} {\bibfnamefont {X.}~\bibnamefont {Li}}, \bibinfo {author}
  {\bibfnamefont {E.~R.}\ \bibnamefont {Warrick}}, \bibinfo {author}
  {\bibfnamefont {M.~J.}\ \bibnamefont {Bell}}, \bibinfo {author}
  {\bibfnamefont {D.~J.}\ \bibnamefont {Haxton}}, \bibinfo {author}
  {\bibfnamefont {C.~W.}\ \bibnamefont {McCurdy}}, \bibinfo {author}
  {\bibfnamefont {D.~M.}\ \bibnamefont {Neumark}}, \ and\ \bibinfo {author}
  {\bibfnamefont {S.~R.}\ \bibnamefont {Leone}},\ }\href {\doibase
  10.1103/PhysRevA.89.023408} {\bibfield  {journal} {\bibinfo  {journal} {Phys.
  Rev. A}\ }\textbf {\bibinfo {volume} {89}},\ \bibinfo {pages} {023408}
  (\bibinfo {year} {2014})}\BibitemShut {NoStop}%
\bibitem [{\citenamefont {Niikura}\ \emph {et~al.}(2003)\citenamefont
  {Niikura}, \citenamefont {Légaré}, \citenamefont {Hasbani}, \citenamefont
  {Ivanov}, \citenamefont {Villeneuve},\ and\ \citenamefont
  {Corkum}}]{Niikura2003}%
  \BibitemOpen
  \bibfield  {author} {\bibinfo {author} {\bibfnamefont {H.}~\bibnamefont
  {Niikura}}, \bibinfo {author} {\bibfnamefont {F.}~\bibnamefont {Légaré}},
  \bibinfo {author} {\bibfnamefont {R.}~\bibnamefont {Hasbani}}, \bibinfo
  {author} {\bibfnamefont {M.~Y.}\ \bibnamefont {Ivanov}}, \bibinfo {author}
  {\bibfnamefont {D.~M.}\ \bibnamefont {Villeneuve}}, \ and\ \bibinfo {author}
  {\bibfnamefont {P.~B.}\ \bibnamefont {Corkum}},\ }\href
  {https://doi.org/10.1038/nature01430} {\bibfield  {journal} {\bibinfo
  {journal} {Nature}\ }\textbf {\bibinfo {volume} {421}},\ \bibinfo {pages}
  {826} (\bibinfo {year} {2003})}\BibitemShut {NoStop}%
\bibitem [{\citenamefont {Kling}\ \emph {et~al.}(2006)\citenamefont {Kling},
  \citenamefont {Siedschlag}, \citenamefont {Verhoef}, \citenamefont {Khan},
  \citenamefont {Schultze}, \citenamefont {Uphues}, \citenamefont {Ni},
  \citenamefont {Uiberacker}, \citenamefont {Drescher}, \citenamefont
  {Krausz},\ and\ \citenamefont {Vrakking}}]{Kling2006}%
  \BibitemOpen
  \bibfield  {author} {\bibinfo {author} {\bibfnamefont {M.~F.}\ \bibnamefont
  {Kling}}, \bibinfo {author} {\bibfnamefont {C.}~\bibnamefont {Siedschlag}},
  \bibinfo {author} {\bibfnamefont {A.~J.}\ \bibnamefont {Verhoef}}, \bibinfo
  {author} {\bibfnamefont {J.~I.}\ \bibnamefont {Khan}}, \bibinfo {author}
  {\bibfnamefont {M.}~\bibnamefont {Schultze}}, \bibinfo {author}
  {\bibfnamefont {T.}~\bibnamefont {Uphues}}, \bibinfo {author} {\bibfnamefont
  {Y.}~\bibnamefont {Ni}}, \bibinfo {author} {\bibfnamefont {M.}~\bibnamefont
  {Uiberacker}}, \bibinfo {author} {\bibfnamefont {M.}~\bibnamefont
  {Drescher}}, \bibinfo {author} {\bibfnamefont {F.}~\bibnamefont {Krausz}}, \
  and\ \bibinfo {author} {\bibfnamefont {M.~J.~J.}\ \bibnamefont {Vrakking}},\
  }\href {\doibase 10.1126/science.1126259} {\bibfield  {journal} {\bibinfo
  {journal} {Science}\ }\textbf {\bibinfo {volume} {312}},\ \bibinfo {pages}
  {246} (\bibinfo {year} {2006})}\BibitemShut {NoStop}%
\bibitem [{\citenamefont {Staudte}\ \emph {et~al.}(2007)\citenamefont
  {Staudte}, \citenamefont {Pavi\ifmmode \check{c}\else
  \v{c}\fi{}i\ifmmode~\acute{c}\else \'{c}\fi{}}, \citenamefont {Chelkowski},
  \citenamefont {Zeidler}, \citenamefont {Meckel}, \citenamefont {Niikura},
  \citenamefont {Sch\"offler}, \citenamefont {Sch\"ossler}, \citenamefont
  {Ulrich}, \citenamefont {Rajeev}, \citenamefont {Weber}, \citenamefont
  {Jahnke}, \citenamefont {Villeneuve}, \citenamefont {Bandrauk}, \citenamefont
  {Cocke}, \citenamefont {Corkum},\ and\ \citenamefont
  {D\"orner}}]{Staudte2007}%
  \BibitemOpen
  \bibfield  {author} {\bibinfo {author} {\bibfnamefont {A.}~\bibnamefont
  {Staudte}}, \bibinfo {author} {\bibfnamefont {D.}~\bibnamefont {Pavi\ifmmode
  \check{c}\else \v{c}\fi{}i\ifmmode~\acute{c}\else \'{c}\fi{}}}, \bibinfo
  {author} {\bibfnamefont {S.}~\bibnamefont {Chelkowski}}, \bibinfo {author}
  {\bibfnamefont {D.}~\bibnamefont {Zeidler}}, \bibinfo {author} {\bibfnamefont
  {M.}~\bibnamefont {Meckel}}, \bibinfo {author} {\bibfnamefont
  {H.}~\bibnamefont {Niikura}}, \bibinfo {author} {\bibfnamefont
  {M.}~\bibnamefont {Sch\"offler}}, \bibinfo {author} {\bibfnamefont
  {S.}~\bibnamefont {Sch\"ossler}}, \bibinfo {author} {\bibfnamefont
  {B.}~\bibnamefont {Ulrich}}, \bibinfo {author} {\bibfnamefont {P.~P.}\
  \bibnamefont {Rajeev}}, \bibinfo {author} {\bibfnamefont {T.}~\bibnamefont
  {Weber}}, \bibinfo {author} {\bibfnamefont {T.}~\bibnamefont {Jahnke}},
  \bibinfo {author} {\bibfnamefont {D.~M.}\ \bibnamefont {Villeneuve}},
  \bibinfo {author} {\bibfnamefont {A.~D.}\ \bibnamefont {Bandrauk}}, \bibinfo
  {author} {\bibfnamefont {C.~L.}\ \bibnamefont {Cocke}}, \bibinfo {author}
  {\bibfnamefont {P.~B.}\ \bibnamefont {Corkum}}, \ and\ \bibinfo {author}
  {\bibfnamefont {R.}~\bibnamefont {D\"orner}},\ }\href {\doibase
  10.1103/PhysRevLett.98.073003} {\bibfield  {journal} {\bibinfo  {journal}
  {Phys. Rev. Lett.}\ }\textbf {\bibinfo {volume} {98}},\ \bibinfo {pages}
  {073003} (\bibinfo {year} {2007})}\BibitemShut {NoStop}%
\bibitem [{\citenamefont {Leone}\ \emph {et~al.}(2014)\citenamefont {Leone},
  \citenamefont {McCurdy}, \citenamefont {Burgdörfer}, \citenamefont
  {Cederbaum}, \citenamefont {Chang}, \citenamefont {Dudovich}, \citenamefont
  {Feist}, \citenamefont {Greene}, \citenamefont {Ivanov}, \citenamefont
  {Kienberger}, \citenamefont {Keller}, \citenamefont {Kling}, \citenamefont
  {Loh}, \citenamefont {Pfeifer}, \citenamefont {Pfeiffer}, \citenamefont
  {Santra}, \citenamefont {Schafer}, \citenamefont {Stolow}, \citenamefont
  {Thumm},\ and\ \citenamefont {Vrakking}}]{Leone2014}%
  \BibitemOpen
  \bibfield  {author} {\bibinfo {author} {\bibfnamefont {S.~R.}\ \bibnamefont
  {Leone}}, \bibinfo {author} {\bibfnamefont {C.~W.}\ \bibnamefont {McCurdy}},
  \bibinfo {author} {\bibfnamefont {J.}~\bibnamefont {Burgdörfer}}, \bibinfo
  {author} {\bibfnamefont {L.~S.}\ \bibnamefont {Cederbaum}}, \bibinfo {author}
  {\bibfnamefont {Z.}~\bibnamefont {Chang}}, \bibinfo {author} {\bibfnamefont
  {N.}~\bibnamefont {Dudovich}}, \bibinfo {author} {\bibfnamefont
  {J.}~\bibnamefont {Feist}}, \bibinfo {author} {\bibfnamefont {C.~H.}\
  \bibnamefont {Greene}}, \bibinfo {author} {\bibfnamefont {M.}~\bibnamefont
  {Ivanov}}, \bibinfo {author} {\bibfnamefont {R.}~\bibnamefont {Kienberger}},
  \bibinfo {author} {\bibfnamefont {U.}~\bibnamefont {Keller}}, \bibinfo
  {author} {\bibfnamefont {M.~F.}\ \bibnamefont {Kling}}, \bibinfo {author}
  {\bibfnamefont {Z.-H.}\ \bibnamefont {Loh}}, \bibinfo {author} {\bibfnamefont
  {T.}~\bibnamefont {Pfeifer}}, \bibinfo {author} {\bibfnamefont {A.~N.}\
  \bibnamefont {Pfeiffer}}, \bibinfo {author} {\bibfnamefont {R.}~\bibnamefont
  {Santra}}, \bibinfo {author} {\bibfnamefont {K.}~\bibnamefont {Schafer}},
  \bibinfo {author} {\bibfnamefont {A.}~\bibnamefont {Stolow}}, \bibinfo
  {author} {\bibfnamefont {U.}~\bibnamefont {Thumm}}, \ and\ \bibinfo {author}
  {\bibfnamefont {M.~J.~J.}\ \bibnamefont {Vrakking}},\ }\href
  {https://doi.org/10.1038/nphoton.2014.48} {\bibfield  {journal} {\bibinfo
  {journal} {Nat. Photon.}\ }\textbf {\bibinfo {volume} {8}},\ \bibinfo {pages}
  {162} (\bibinfo {year} {2014})}\BibitemShut {NoStop}%
\bibitem [{\citenamefont {F{\"o}rg}\ \emph {et~al.}(2016)\citenamefont
  {F{\"o}rg}, \citenamefont {Sch{\"o}tz}, \citenamefont {S{\"u}{\ss}mann},
  \citenamefont {F{\"o}rster}, \citenamefont {Kr{\"u}ger}, \citenamefont {Ahn},
  \citenamefont {Okell}, \citenamefont {Wintersperger}, \citenamefont
  {Zherebtsov}, \citenamefont {Guggenmos}, \citenamefont {Pervak},
  \citenamefont {Kessel}, \citenamefont {Trushin}, \citenamefont {Azzeer},
  \citenamefont {Stockman}, \citenamefont {Kim}, \citenamefont {Krausz},
  \citenamefont {Hommelhoff},\ and\ \citenamefont {Kling}}]{Forg2016}%
  \BibitemOpen
  \bibfield  {author} {\bibinfo {author} {\bibfnamefont {B.}~\bibnamefont
  {F{\"o}rg}}, \bibinfo {author} {\bibfnamefont {J.}~\bibnamefont
  {Sch{\"o}tz}}, \bibinfo {author} {\bibfnamefont {F.}~\bibnamefont
  {S{\"u}{\ss}mann}}, \bibinfo {author} {\bibfnamefont {M.}~\bibnamefont
  {F{\"o}rster}}, \bibinfo {author} {\bibfnamefont {M.}~\bibnamefont
  {Kr{\"u}ger}}, \bibinfo {author} {\bibfnamefont {B.}~\bibnamefont {Ahn}},
  \bibinfo {author} {\bibfnamefont {W.~A.}\ \bibnamefont {Okell}}, \bibinfo
  {author} {\bibfnamefont {K.}~\bibnamefont {Wintersperger}}, \bibinfo {author}
  {\bibfnamefont {S.}~\bibnamefont {Zherebtsov}}, \bibinfo {author}
  {\bibfnamefont {A.}~\bibnamefont {Guggenmos}}, \bibinfo {author}
  {\bibfnamefont {V.}~\bibnamefont {Pervak}}, \bibinfo {author} {\bibfnamefont
  {A.}~\bibnamefont {Kessel}}, \bibinfo {author} {\bibfnamefont {S.~A.}\
  \bibnamefont {Trushin}}, \bibinfo {author} {\bibfnamefont {A.~M.}\
  \bibnamefont {Azzeer}}, \bibinfo {author} {\bibfnamefont {M.~I.}\
  \bibnamefont {Stockman}}, \bibinfo {author} {\bibfnamefont {D.}~\bibnamefont
  {Kim}}, \bibinfo {author} {\bibfnamefont {F.}~\bibnamefont {Krausz}},
  \bibinfo {author} {\bibfnamefont {P.}~\bibnamefont {Hommelhoff}}, \ and\
  \bibinfo {author} {\bibfnamefont {M.~F.}\ \bibnamefont {Kling}},\ }\href
  {https://doi.org/10.1038/ncomms11717} {\bibfield  {journal} {\bibinfo
  {journal} {Nat. Commun.}\ }\textbf {\bibinfo {volume} {7}},\ \bibinfo {pages}
  {11717} (\bibinfo {year} {2016})}\BibitemShut {NoStop}%
\bibitem [{\citenamefont {Li}\ \emph {et~al.}(2016)\citenamefont {Li},
  \citenamefont {Saydanzad},\ and\ \citenamefont {Thumm}}]{Li2016}%
  \BibitemOpen
  \bibfield  {author} {\bibinfo {author} {\bibfnamefont {J.}~\bibnamefont
  {Li}}, \bibinfo {author} {\bibfnamefont {E.}~\bibnamefont {Saydanzad}}, \
  and\ \bibinfo {author} {\bibfnamefont {U.}~\bibnamefont {Thumm}},\ }\href
  {\doibase 10.1103/PhysRevA.94.051401} {\bibfield  {journal} {\bibinfo
  {journal} {Phys. Rev. A}\ }\textbf {\bibinfo {volume} {94}},\ \bibinfo
  {pages} {051401} (\bibinfo {year} {2016})}\BibitemShut {NoStop}%
\bibitem [{\citenamefont {Seiffert}\ \emph {et~al.}(2017)\citenamefont
  {Seiffert}, \citenamefont {Liu}, \citenamefont {Zherebtsov}, \citenamefont
  {Trabattoni}, \citenamefont {Rupp}, \citenamefont {Castrovilli},
  \citenamefont {Galli}, \citenamefont {S\"{u}{\ss}mann}, \citenamefont
  {Wintersperger}, \citenamefont {Stierle}, \citenamefont {Sansone},
  \citenamefont {Poletto}, \citenamefont {Frassetto}, \citenamefont {Halfpap},
  \citenamefont {Mondes}, \citenamefont {Graf}, \citenamefont {R\"{u}hl},
  \citenamefont {Krausz}, \citenamefont {Nisoli}, \citenamefont {Fennel},
  \citenamefont {Calegari},\ and\ \citenamefont {Kling}}]{Seiffert2017}%
  \BibitemOpen
  \bibfield  {author} {\bibinfo {author} {\bibfnamefont {L.}~\bibnamefont
  {Seiffert}}, \bibinfo {author} {\bibfnamefont {Q.}~\bibnamefont {Liu}},
  \bibinfo {author} {\bibfnamefont {S.}~\bibnamefont {Zherebtsov}}, \bibinfo
  {author} {\bibfnamefont {A.}~\bibnamefont {Trabattoni}}, \bibinfo {author}
  {\bibfnamefont {P.}~\bibnamefont {Rupp}}, \bibinfo {author} {\bibfnamefont
  {M.~C.}\ \bibnamefont {Castrovilli}}, \bibinfo {author} {\bibfnamefont
  {M.}~\bibnamefont {Galli}}, \bibinfo {author} {\bibfnamefont
  {F.}~\bibnamefont {S\"{u}{\ss}mann}}, \bibinfo {author} {\bibfnamefont
  {K.}~\bibnamefont {Wintersperger}}, \bibinfo {author} {\bibfnamefont
  {J.}~\bibnamefont {Stierle}}, \bibinfo {author} {\bibfnamefont
  {G.}~\bibnamefont {Sansone}}, \bibinfo {author} {\bibfnamefont
  {L.}~\bibnamefont {Poletto}}, \bibinfo {author} {\bibfnamefont
  {F.}~\bibnamefont {Frassetto}}, \bibinfo {author} {\bibfnamefont
  {I.}~\bibnamefont {Halfpap}}, \bibinfo {author} {\bibfnamefont
  {V.}~\bibnamefont {Mondes}}, \bibinfo {author} {\bibfnamefont
  {C.}~\bibnamefont {Graf}}, \bibinfo {author} {\bibfnamefont {E.}~\bibnamefont
  {R\"{u}hl}}, \bibinfo {author} {\bibfnamefont {F.}~\bibnamefont {Krausz}},
  \bibinfo {author} {\bibfnamefont {M.}~\bibnamefont {Nisoli}}, \bibinfo
  {author} {\bibfnamefont {T.}~\bibnamefont {Fennel}}, \bibinfo {author}
  {\bibfnamefont {F.}~\bibnamefont {Calegari}}, \ and\ \bibinfo {author}
  {\bibfnamefont {M.}~\bibnamefont {Kling}},\ }\href
  {https://doi.org/10.1038/nphys4129} {\bibfield  {journal} {\bibinfo
  {journal} {Nat. Phys.}\ }\textbf {\bibinfo {volume} {13}},\ \bibinfo {pages}
  {766} (\bibinfo {year} {2017})}\BibitemShut {NoStop}%
\bibitem [{\citenamefont {Sch{\"o}tz}\ \emph {et~al.}(2017)\citenamefont
  {Sch{\"o}tz}, \citenamefont {F{\"o}rg}, \citenamefont {F{\"o}rster},
  \citenamefont {Okell}, \citenamefont {Stockman}, \citenamefont {Krausz},
  \citenamefont {Hommelhoff},\ and\ \citenamefont {Kling}}]{Schoetz2017}%
  \BibitemOpen
  \bibfield  {author} {\bibinfo {author} {\bibfnamefont {J.}~\bibnamefont
  {Sch{\"o}tz}}, \bibinfo {author} {\bibfnamefont {B.}~\bibnamefont
  {F{\"o}rg}}, \bibinfo {author} {\bibfnamefont {M.}~\bibnamefont
  {F{\"o}rster}}, \bibinfo {author} {\bibfnamefont {W.~A.}\ \bibnamefont
  {Okell}}, \bibinfo {author} {\bibfnamefont {M.~I.}\ \bibnamefont {Stockman}},
  \bibinfo {author} {\bibfnamefont {F.}~\bibnamefont {Krausz}}, \bibinfo
  {author} {\bibfnamefont {P.}~\bibnamefont {Hommelhoff}}, \ and\ \bibinfo
  {author} {\bibfnamefont {M.~F.}\ \bibnamefont {Kling}},\ }\href@noop {}
  {\bibfield  {journal} {\bibinfo  {journal} {IEEE Journal of Selected Topics
  in Quantum Electronics}\ }\textbf {\bibinfo {volume} {23}},\ \bibinfo {pages}
  {77} (\bibinfo {year} {2017})}\BibitemShut {NoStop}%
\bibitem [{\citenamefont {Saydanzad}\ \emph {et~al.}(2017)\citenamefont
  {Saydanzad}, \citenamefont {Li},\ and\ \citenamefont
  {Thumm}}]{Saydanzad2017}%
  \BibitemOpen
  \bibfield  {author} {\bibinfo {author} {\bibfnamefont {E.}~\bibnamefont
  {Saydanzad}}, \bibinfo {author} {\bibfnamefont {J.}~\bibnamefont {Li}}, \
  and\ \bibinfo {author} {\bibfnamefont {U.}~\bibnamefont {Thumm}},\ }\href
  {\doibase 10.1103/PhysRevA.95.053406} {\bibfield  {journal} {\bibinfo
  {journal} {Phys. Rev. A}\ }\textbf {\bibinfo {volume} {95}},\ \bibinfo
  {pages} {053406} (\bibinfo {year} {2017})}\BibitemShut {NoStop}%
\bibitem [{\citenamefont {Li}\ \emph {et~al.}(2017)\citenamefont {Li},
  \citenamefont {Saydanzad},\ and\ \citenamefont {Thumm}}]{Li2017}%
  \BibitemOpen
  \bibfield  {author} {\bibinfo {author} {\bibfnamefont {J.}~\bibnamefont
  {Li}}, \bibinfo {author} {\bibfnamefont {E.}~\bibnamefont {Saydanzad}}, \
  and\ \bibinfo {author} {\bibfnamefont {U.}~\bibnamefont {Thumm}},\ }\href
  {\doibase 10.1103/PhysRevA.95.043423} {\bibfield  {journal} {\bibinfo
  {journal} {Phys. Rev. A}\ }\textbf {\bibinfo {volume} {95}},\ \bibinfo
  {pages} {043423} (\bibinfo {year} {2017})}\BibitemShut {NoStop}%
\bibitem [{\citenamefont {Saydanzad}\ \emph {et~al.}(2018)\citenamefont
  {Saydanzad}, \citenamefont {Li},\ and\ \citenamefont
  {Thumm}}]{Saydanzad2018}%
  \BibitemOpen
  \bibfield  {author} {\bibinfo {author} {\bibfnamefont {E.}~\bibnamefont
  {Saydanzad}}, \bibinfo {author} {\bibfnamefont {J.}~\bibnamefont {Li}}, \
  and\ \bibinfo {author} {\bibfnamefont {U.}~\bibnamefont {Thumm}},\ }\href
  {\doibase 10.1103/PhysRevA.98.063422} {\bibfield  {journal} {\bibinfo
  {journal} {Phys. Rev. A}\ }\textbf {\bibinfo {volume} {98}},\ \bibinfo
  {pages} {063422} (\bibinfo {year} {2018})}\BibitemShut {NoStop}%
\bibitem [{\citenamefont {Li}\ \emph {et~al.}(2018)\citenamefont {Li},
  \citenamefont {Saydanzad},\ and\ \citenamefont {Thumm}}]{Li2018PRL}%
  \BibitemOpen
  \bibfield  {author} {\bibinfo {author} {\bibfnamefont {J.}~\bibnamefont
  {Li}}, \bibinfo {author} {\bibfnamefont {E.}~\bibnamefont {Saydanzad}}, \
  and\ \bibinfo {author} {\bibfnamefont {U.}~\bibnamefont {Thumm}},\ }\href
  {\doibase 10.1103/PhysRevLett.120.223903} {\bibfield  {journal} {\bibinfo
  {journal} {Phys. Rev. Lett.}\ }\textbf {\bibinfo {volume} {120}},\ \bibinfo
  {pages} {223903} (\bibinfo {year} {2018})}\BibitemShut {NoStop}%
\bibitem [{\citenamefont {Lucchini}\ \emph {et~al.}(2015)\citenamefont
  {Lucchini}, \citenamefont {Ludwig}, \citenamefont {Kasmi}, \citenamefont
  {Gallmann},\ and\ \citenamefont {Keller}}]{Lucchini2015}%
  \BibitemOpen
  \bibfield  {author} {\bibinfo {author} {\bibfnamefont {M.}~\bibnamefont
  {Lucchini}}, \bibinfo {author} {\bibfnamefont {A.}~\bibnamefont {Ludwig}},
  \bibinfo {author} {\bibfnamefont {L.}~\bibnamefont {Kasmi}}, \bibinfo
  {author} {\bibfnamefont {L.}~\bibnamefont {Gallmann}}, \ and\ \bibinfo
  {author} {\bibfnamefont {U.}~\bibnamefont {Keller}},\ }\href {\doibase
  10.1364/OE.23.008867} {\bibfield  {journal} {\bibinfo  {journal} {Opt.
  Express}\ }\textbf {\bibinfo {volume} {23}},\ \bibinfo {pages} {8867}
  (\bibinfo {year} {2015})}\BibitemShut {NoStop}%
\bibitem [{\citenamefont {Chen}\ \emph {et~al.}(2017)\citenamefont {Chen},
  \citenamefont {Tao}, \citenamefont {Carr}, \citenamefont {Matyba},
  \citenamefont {SzilvÃ¡si}, \citenamefont {Emmerich}, \citenamefont {Piecuch},
  \citenamefont {Keller}, \citenamefont {Zusin}, \citenamefont {Eich},
  \citenamefont {Rollinger}, \citenamefont {You}, \citenamefont {Mathias},
  \citenamefont {Thumm}, \citenamefont {Mavrikakis}, \citenamefont
  {Aeschlimann}, \citenamefont {Oppeneer}, \citenamefont {Kapteyn},\ and\
  \citenamefont {Murnane}}]{Chen2017}%
  \BibitemOpen
  \bibfield  {author} {\bibinfo {author} {\bibfnamefont {C.}~\bibnamefont
  {Chen}}, \bibinfo {author} {\bibfnamefont {Z.}~\bibnamefont {Tao}}, \bibinfo
  {author} {\bibfnamefont {A.}~\bibnamefont {Carr}}, \bibinfo {author}
  {\bibfnamefont {P.}~\bibnamefont {Matyba}}, \bibinfo {author} {\bibfnamefont
  {T.}~\bibnamefont {SzilvÃ¡si}}, \bibinfo {author} {\bibfnamefont
  {S.}~\bibnamefont {Emmerich}}, \bibinfo {author} {\bibfnamefont
  {M.}~\bibnamefont {Piecuch}}, \bibinfo {author} {\bibfnamefont
  {M.}~\bibnamefont {Keller}}, \bibinfo {author} {\bibfnamefont
  {D.}~\bibnamefont {Zusin}}, \bibinfo {author} {\bibfnamefont
  {S.}~\bibnamefont {Eich}}, \bibinfo {author} {\bibfnamefont {M.}~\bibnamefont
  {Rollinger}}, \bibinfo {author} {\bibfnamefont {W.}~\bibnamefont {You}},
  \bibinfo {author} {\bibfnamefont {S.}~\bibnamefont {Mathias}}, \bibinfo
  {author} {\bibfnamefont {U.}~\bibnamefont {Thumm}}, \bibinfo {author}
  {\bibfnamefont {M.}~\bibnamefont {Mavrikakis}}, \bibinfo {author}
  {\bibfnamefont {M.}~\bibnamefont {Aeschlimann}}, \bibinfo {author}
  {\bibfnamefont {P.~M.}\ \bibnamefont {Oppeneer}}, \bibinfo {author}
  {\bibfnamefont {H.}~\bibnamefont {Kapteyn}}, \ and\ \bibinfo {author}
  {\bibfnamefont {M.}~\bibnamefont {Murnane}},\ }\href {\doibase
  10.1073/pnas.1706466114} {\bibfield  {journal} {\bibinfo  {journal} {Proc.
  Natl. Acad. Sci. USA}\ }\textbf {\bibinfo {volume} {114}},\ \bibinfo {pages}
  {E5300} (\bibinfo {year} {2017})}\BibitemShut {NoStop}%
\bibitem [{\citenamefont {Neppl}\ \emph
  {et~al.}(2015{\natexlab{a}})\citenamefont {Neppl}, \citenamefont
  {Ernstorfer}, \citenamefont {Cavalieri}, \citenamefont {Lemell},
  \citenamefont {Wachter}, \citenamefont {Magerl}, \citenamefont
  {Bothschafter}, \citenamefont {Jobst}, \citenamefont {Hofstetter},
  \citenamefont {Kleineberg}, \citenamefont {Barth}, \citenamefont {Menzel},
  \citenamefont {Burgd\"{o}rfer}, \citenamefont {Feulner}, \citenamefont
  {Krausz},\ and\ \citenamefont {Kienberger}}]{Neppl2015}%
  \BibitemOpen
  \bibfield  {author} {\bibinfo {author} {\bibfnamefont {S.}~\bibnamefont
  {Neppl}}, \bibinfo {author} {\bibfnamefont {R.}~\bibnamefont {Ernstorfer}},
  \bibinfo {author} {\bibfnamefont {A.~L.}\ \bibnamefont {Cavalieri}}, \bibinfo
  {author} {\bibfnamefont {C.}~\bibnamefont {Lemell}}, \bibinfo {author}
  {\bibfnamefont {G.}~\bibnamefont {Wachter}}, \bibinfo {author} {\bibfnamefont
  {E.}~\bibnamefont {Magerl}}, \bibinfo {author} {\bibfnamefont {E.~M.}\
  \bibnamefont {Bothschafter}}, \bibinfo {author} {\bibfnamefont
  {M.}~\bibnamefont {Jobst}}, \bibinfo {author} {\bibfnamefont
  {M.}~\bibnamefont {Hofstetter}}, \bibinfo {author} {\bibfnamefont
  {U.}~\bibnamefont {Kleineberg}}, \bibinfo {author} {\bibfnamefont {J.~V.}\
  \bibnamefont {Barth}}, \bibinfo {author} {\bibfnamefont {D.}~\bibnamefont
  {Menzel}}, \bibinfo {author} {\bibfnamefont {J.}~\bibnamefont
  {Burgd\"{o}rfer}}, \bibinfo {author} {\bibfnamefont {P.}~\bibnamefont
  {Feulner}}, \bibinfo {author} {\bibfnamefont {F.}~\bibnamefont {Krausz}}, \
  and\ \bibinfo {author} {\bibfnamefont {R.}~\bibnamefont {Kienberger}},\
  }\href {\doibase 10.1038/nature14094} {\bibfield  {journal} {\bibinfo
  {journal} {Nature}\ }\textbf {\bibinfo {volume} {517}},\ \bibinfo {pages}
  {342} (\bibinfo {year} {2015}{\natexlab{a}})}\BibitemShut {NoStop}%
\bibitem [{\citenamefont {Tao}\ \emph {et~al.}(2016)\citenamefont {Tao},
  \citenamefont {Chen}, \citenamefont {Szilv{\'a}si}, \citenamefont {Keller},
  \citenamefont {Mavrikakis}, \citenamefont {Kapteyn},\ and\ \citenamefont
  {Murnane}}]{Tao2016}%
  \BibitemOpen
  \bibfield  {author} {\bibinfo {author} {\bibfnamefont {Z.}~\bibnamefont
  {Tao}}, \bibinfo {author} {\bibfnamefont {C.}~\bibnamefont {Chen}}, \bibinfo
  {author} {\bibfnamefont {T.}~\bibnamefont {Szilv{\'a}si}}, \bibinfo {author}
  {\bibfnamefont {M.}~\bibnamefont {Keller}}, \bibinfo {author} {\bibfnamefont
  {M.}~\bibnamefont {Mavrikakis}}, \bibinfo {author} {\bibfnamefont
  {H.}~\bibnamefont {Kapteyn}}, \ and\ \bibinfo {author} {\bibfnamefont
  {M.}~\bibnamefont {Murnane}},\ }\href {\doibase 10.1126/science.aaf6793}
  {\bibfield  {journal} {\bibinfo  {journal} {Science}\ }\textbf {\bibinfo
  {volume} {353}},\ \bibinfo {pages} {62} (\bibinfo {year} {2016})}\BibitemShut
  {NoStop}%
\bibitem [{\citenamefont {Siek}\ \emph {et~al.}(2017)\citenamefont {Siek},
  \citenamefont {Neb}, \citenamefont {Bartz}, \citenamefont {Hensen},
  \citenamefont {Str{\"u}ber}, \citenamefont {Fiechter}, \citenamefont
  {Torrent-Sucarrat}, \citenamefont {Silkin}, \citenamefont {Krasovskii},
  \citenamefont {Kabachnik}, \citenamefont {Fritzsche}, \citenamefont
  {Mui{\~n}o}, \citenamefont {Echenique}, \citenamefont {Kazansky},
  \citenamefont {M{\"u}ller}, \citenamefont {Pfeiffer},\ and\ \citenamefont
  {Heinzmann}}]{Heinzmann2017}%
  \BibitemOpen
  \bibfield  {author} {\bibinfo {author} {\bibfnamefont {F.}~\bibnamefont
  {Siek}}, \bibinfo {author} {\bibfnamefont {S.}~\bibnamefont {Neb}}, \bibinfo
  {author} {\bibfnamefont {P.}~\bibnamefont {Bartz}}, \bibinfo {author}
  {\bibfnamefont {M.}~\bibnamefont {Hensen}}, \bibinfo {author} {\bibfnamefont
  {C.}~\bibnamefont {Str{\"u}ber}}, \bibinfo {author} {\bibfnamefont
  {S.}~\bibnamefont {Fiechter}}, \bibinfo {author} {\bibfnamefont
  {M.}~\bibnamefont {Torrent-Sucarrat}}, \bibinfo {author} {\bibfnamefont
  {V.~M.}\ \bibnamefont {Silkin}}, \bibinfo {author} {\bibfnamefont {E.~E.}\
  \bibnamefont {Krasovskii}}, \bibinfo {author} {\bibfnamefont {N.~M.}\
  \bibnamefont {Kabachnik}}, \bibinfo {author} {\bibfnamefont {S.}~\bibnamefont
  {Fritzsche}}, \bibinfo {author} {\bibfnamefont {R.~D.}\ \bibnamefont
  {Mui{\~n}o}}, \bibinfo {author} {\bibfnamefont {P.~M.}\ \bibnamefont
  {Echenique}}, \bibinfo {author} {\bibfnamefont {A.~K.}\ \bibnamefont
  {Kazansky}}, \bibinfo {author} {\bibfnamefont {N.}~\bibnamefont
  {M{\"u}ller}}, \bibinfo {author} {\bibfnamefont {W.}~\bibnamefont
  {Pfeiffer}}, \ and\ \bibinfo {author} {\bibfnamefont {U.}~\bibnamefont
  {Heinzmann}},\ }\href {\doibase 10.1126/science.aam9598} {\bibfield
  {journal} {\bibinfo  {journal} {Science}\ }\textbf {\bibinfo {volume}
  {357}},\ \bibinfo {pages} {1274} (\bibinfo {year} {2017})}\BibitemShut
  {NoStop}%
\bibitem [{\citenamefont {Kasmi}\ \emph {et~al.}(2017)\citenamefont {Kasmi},
  \citenamefont {Lucchini}, \citenamefont {Castiglioni}, \citenamefont
  {Kliuiev}, \citenamefont {Osterwalder}, \citenamefont {Hengsberger},
  \citenamefont {Gallmann}, \citenamefont {Kr\"{u}ger},\ and\ \citenamefont
  {Keller}}]{Kasmi2017}%
  \BibitemOpen
  \bibfield  {author} {\bibinfo {author} {\bibfnamefont {L.}~\bibnamefont
  {Kasmi}}, \bibinfo {author} {\bibfnamefont {M.}~\bibnamefont {Lucchini}},
  \bibinfo {author} {\bibfnamefont {L.}~\bibnamefont {Castiglioni}}, \bibinfo
  {author} {\bibfnamefont {P.}~\bibnamefont {Kliuiev}}, \bibinfo {author}
  {\bibfnamefont {J.}~\bibnamefont {Osterwalder}}, \bibinfo {author}
  {\bibfnamefont {M.}~\bibnamefont {Hengsberger}}, \bibinfo {author}
  {\bibfnamefont {L.}~\bibnamefont {Gallmann}}, \bibinfo {author}
  {\bibfnamefont {P.}~\bibnamefont {Kr\"{u}ger}}, \ and\ \bibinfo {author}
  {\bibfnamefont {U.}~\bibnamefont {Keller}},\ }\href {\doibase
  10.1364/OPTICA.4.001492} {\bibfield  {journal} {\bibinfo  {journal} {Optica}\
  }\textbf {\bibinfo {volume} {4}},\ \bibinfo {pages} {1492} (\bibinfo {year}
  {2017})}\BibitemShut {NoStop}%
\bibitem [{\citenamefont {Ambrosio}\ and\ \citenamefont
  {Thumm}(2018)}]{Ambrosio2018}%
  \BibitemOpen
  \bibfield  {author} {\bibinfo {author} {\bibfnamefont {M.~J.}\ \bibnamefont
  {Ambrosio}}\ and\ \bibinfo {author} {\bibfnamefont {U.}~\bibnamefont
  {Thumm}},\ }\href {\doibase 10.1103/PhysRevA.97.043431} {\bibfield  {journal}
  {\bibinfo  {journal} {Phys. Rev. A}\ }\textbf {\bibinfo {volume} {97}},\
  \bibinfo {pages} {043431} (\bibinfo {year} {2018})}\BibitemShut {NoStop}%
\bibitem [{\citenamefont {Ambrosio}\ and\ \citenamefont
  {Thumm}(2019)}]{Ambrosio2019}%
  \BibitemOpen
  \bibfield  {author} {\bibinfo {author} {\bibfnamefont {M.~J.}\ \bibnamefont
  {Ambrosio}}\ and\ \bibinfo {author} {\bibfnamefont {U.}~\bibnamefont
  {Thumm}},\ }\href {\doibase 10.1103/PhysRevA.100.043412} {\bibfield
  {journal} {\bibinfo  {journal} {Phys. Rev. A}\ }\textbf {\bibinfo {volume}
  {100}},\ \bibinfo {pages} {043412} (\bibinfo {year} {2019})}\BibitemShut
  {NoStop}%
\bibitem [{\citenamefont {Ossiander}\ \emph {et~al.}(2018)\citenamefont
  {Ossiander}, \citenamefont {Riemensberger}, \citenamefont {Neppl},
  \citenamefont {Mittermair}, \citenamefont {Sch{\"a}ffer}, \citenamefont
  {Duensing}, \citenamefont {Wagner}, \citenamefont {Heider}, \citenamefont
  {Wurzer}, \citenamefont {Gerl}, \citenamefont {Schnitzenbaumer},
  \citenamefont {Barth}, \citenamefont {Libisch}, \citenamefont {Lemell},
  \citenamefont {Burgd{\"o}rfer}, \citenamefont {Feulner},\ and\ \citenamefont
  {Kienberger}}]{Ossiander2018}%
  \BibitemOpen
  \bibfield  {author} {\bibinfo {author} {\bibfnamefont {M.}~\bibnamefont
  {Ossiander}}, \bibinfo {author} {\bibfnamefont {J.}~\bibnamefont
  {Riemensberger}}, \bibinfo {author} {\bibfnamefont {S.}~\bibnamefont
  {Neppl}}, \bibinfo {author} {\bibfnamefont {M.}~\bibnamefont {Mittermair}},
  \bibinfo {author} {\bibfnamefont {M.}~\bibnamefont {Sch{\"a}ffer}}, \bibinfo
  {author} {\bibfnamefont {A.}~\bibnamefont {Duensing}}, \bibinfo {author}
  {\bibfnamefont {M.~S.}\ \bibnamefont {Wagner}}, \bibinfo {author}
  {\bibfnamefont {R.}~\bibnamefont {Heider}}, \bibinfo {author} {\bibfnamefont
  {M.}~\bibnamefont {Wurzer}}, \bibinfo {author} {\bibfnamefont
  {M.}~\bibnamefont {Gerl}}, \bibinfo {author} {\bibfnamefont {M.}~\bibnamefont
  {Schnitzenbaumer}}, \bibinfo {author} {\bibfnamefont {J.~V.}\ \bibnamefont
  {Barth}}, \bibinfo {author} {\bibfnamefont {F.}~\bibnamefont {Libisch}},
  \bibinfo {author} {\bibfnamefont {C.}~\bibnamefont {Lemell}}, \bibinfo
  {author} {\bibfnamefont {J.}~\bibnamefont {Burgd{\"o}rfer}}, \bibinfo
  {author} {\bibfnamefont {P.}~\bibnamefont {Feulner}}, \ and\ \bibinfo
  {author} {\bibfnamefont {R.}~\bibnamefont {Kienberger}},\ }\href
  {https://doi.org/10.1038/s41586-018-0503-6} {\bibfield  {journal} {\bibinfo
  {journal} {Nature}\ }\textbf {\bibinfo {volume} {561}},\ \bibinfo {pages}
  {374} (\bibinfo {year} {2018})}\BibitemShut {NoStop}%
\bibitem [{\citenamefont {Zhang}\ and\ \citenamefont
  {Thumm}(2011)}]{Zhang2011}%
  \BibitemOpen
  \bibfield  {author} {\bibinfo {author} {\bibfnamefont {C.-H.}\ \bibnamefont
  {Zhang}}\ and\ \bibinfo {author} {\bibfnamefont {U.}~\bibnamefont {Thumm}},\
  }\href {\doibase 10.1103/PhysRevA.84.063403} {\bibfield  {journal} {\bibinfo
  {journal} {Phys. Rev. A}\ }\textbf {\bibinfo {volume} {84}},\ \bibinfo
  {pages} {063403} (\bibinfo {year} {2011})}\BibitemShut {NoStop}%
\bibitem [{\citenamefont {Chew}\ \emph {et~al.}(2012)\citenamefont {Chew},
  \citenamefont {S{\"u}ßmann}, \citenamefont {Sp{\"a}th}, \citenamefont
  {Wirth}, \citenamefont {Schmidt}, \citenamefont {Zherebtsov}, \citenamefont
  {Guggenmos}, \citenamefont {Oelsner}, \citenamefont {Weber}, \citenamefont
  {Kapaldo}, \citenamefont {Gliserin}, \citenamefont {Stockman}, \citenamefont
  {Kling},\ and\ \citenamefont {Kleineberg}}]{Chew2012}%
  \BibitemOpen
  \bibfield  {author} {\bibinfo {author} {\bibfnamefont {S.~H.}\ \bibnamefont
  {Chew}}, \bibinfo {author} {\bibfnamefont {F.}~\bibnamefont {S{\"u}ßmann}},
  \bibinfo {author} {\bibfnamefont {C.}~\bibnamefont {Sp{\"a}th}}, \bibinfo
  {author} {\bibfnamefont {A.}~\bibnamefont {Wirth}}, \bibinfo {author}
  {\bibfnamefont {J.}~\bibnamefont {Schmidt}}, \bibinfo {author} {\bibfnamefont
  {S.}~\bibnamefont {Zherebtsov}}, \bibinfo {author} {\bibfnamefont
  {A.}~\bibnamefont {Guggenmos}}, \bibinfo {author} {\bibfnamefont
  {A.}~\bibnamefont {Oelsner}}, \bibinfo {author} {\bibfnamefont
  {N.}~\bibnamefont {Weber}}, \bibinfo {author} {\bibfnamefont
  {J.}~\bibnamefont {Kapaldo}}, \bibinfo {author} {\bibfnamefont
  {A.}~\bibnamefont {Gliserin}}, \bibinfo {author} {\bibfnamefont {M.~I.}\
  \bibnamefont {Stockman}}, \bibinfo {author} {\bibfnamefont {M.~F.}\
  \bibnamefont {Kling}}, \ and\ \bibinfo {author} {\bibfnamefont
  {U.}~\bibnamefont {Kleineberg}},\ }\href {\doibase 10.1063/1.3670324}
  {\bibfield  {journal} {\bibinfo  {journal} {Appl. Phys. Lett.}\ }\textbf
  {\bibinfo {volume} {100}},\ \bibinfo {pages} {051904} (\bibinfo {year}
  {2012})}\BibitemShut {NoStop}%
\bibitem [{\citenamefont {Lupetti}\ \emph {et~al.}(2014)\citenamefont
  {Lupetti}, \citenamefont {Hengster}, \citenamefont {Uphues},\ and\
  \citenamefont {Scrinzi}}]{Lupetti2014}%
  \BibitemOpen
  \bibfield  {author} {\bibinfo {author} {\bibfnamefont {M.}~\bibnamefont
  {Lupetti}}, \bibinfo {author} {\bibfnamefont {J.}~\bibnamefont {Hengster}},
  \bibinfo {author} {\bibfnamefont {T.}~\bibnamefont {Uphues}}, \ and\ \bibinfo
  {author} {\bibfnamefont {A.}~\bibnamefont {Scrinzi}},\ }\href {\doibase
  10.1103/PhysRevLett.113.113903} {\bibfield  {journal} {\bibinfo  {journal}
  {Phys. Rev. Lett.}\ }\textbf {\bibinfo {volume} {113}},\ \bibinfo {pages}
  {113903} (\bibinfo {year} {2014})}\BibitemShut {NoStop}%
\bibitem [{\citenamefont {Lemke}\ \emph {et~al.}(2013)\citenamefont {Lemke},
  \citenamefont {Schneider}, \citenamefont {Lei{\ss}ner}, \citenamefont
  {Bayer}, \citenamefont {Radke}, \citenamefont {Fischer}, \citenamefont
  {Melchior}, \citenamefont {Evlyukhin}, \citenamefont {Chichkov},
  \citenamefont {Reinhardt}, \citenamefont {Bauer},\ and\ \citenamefont
  {Aeschlimann}}]{Lemke2013}%
  \BibitemOpen
  \bibfield  {author} {\bibinfo {author} {\bibfnamefont {C.}~\bibnamefont
  {Lemke}}, \bibinfo {author} {\bibfnamefont {C.}~\bibnamefont {Schneider}},
  \bibinfo {author} {\bibfnamefont {T.}~\bibnamefont {Lei{\ss}ner}}, \bibinfo
  {author} {\bibfnamefont {D.}~\bibnamefont {Bayer}}, \bibinfo {author}
  {\bibfnamefont {J.~W.}\ \bibnamefont {Radke}}, \bibinfo {author}
  {\bibfnamefont {A.}~\bibnamefont {Fischer}}, \bibinfo {author} {\bibfnamefont
  {P.}~\bibnamefont {Melchior}}, \bibinfo {author} {\bibfnamefont {A.~B.}\
  \bibnamefont {Evlyukhin}}, \bibinfo {author} {\bibfnamefont {B.~N.}\
  \bibnamefont {Chichkov}}, \bibinfo {author} {\bibfnamefont {C.}~\bibnamefont
  {Reinhardt}}, \bibinfo {author} {\bibfnamefont {M.}~\bibnamefont {Bauer}}, \
  and\ \bibinfo {author} {\bibfnamefont {M.}~\bibnamefont {Aeschlimann}},\
  }\href@noop {} {\bibfield  {journal} {\bibinfo  {journal} {Nano Letters}\
  }\textbf {\bibinfo {volume} {13}},\ \bibinfo {pages} {1053} (\bibinfo {year}
  {2013})}\BibitemShut {NoStop}%
\bibitem [{\citenamefont {Liao}\ and\ \citenamefont {Thumm}(2015)}]{Liao2015}%
  \BibitemOpen
  \bibfield  {author} {\bibinfo {author} {\bibfnamefont {Q.}~\bibnamefont
  {Liao}}\ and\ \bibinfo {author} {\bibfnamefont {U.}~\bibnamefont {Thumm}},\
  }\href {\doibase 10.1103/PhysRevA.92.031401} {\bibfield  {journal} {\bibinfo
  {journal} {Phys. Rev. A}\ }\textbf {\bibinfo {volume} {92}},\ \bibinfo
  {pages} {031401} (\bibinfo {year} {2015})}\BibitemShut {NoStop}%
\bibitem [{\citenamefont {Schlather}\ \emph {et~al.}(2017)\citenamefont
  {Schlather}, \citenamefont {Manjavacas}, \citenamefont {Lauchner},
  \citenamefont {Marangoni}, \citenamefont {DeSantis}, \citenamefont
  {Nordlander},\ and\ \citenamefont {Halas}}]{Schlather2017}%
  \BibitemOpen
  \bibfield  {author} {\bibinfo {author} {\bibfnamefont {A.~E.}\ \bibnamefont
  {Schlather}}, \bibinfo {author} {\bibfnamefont {A.}~\bibnamefont
  {Manjavacas}}, \bibinfo {author} {\bibfnamefont {A.}~\bibnamefont
  {Lauchner}}, \bibinfo {author} {\bibfnamefont {V.~S.}\ \bibnamefont
  {Marangoni}}, \bibinfo {author} {\bibfnamefont {C.~J.}\ \bibnamefont
  {DeSantis}}, \bibinfo {author} {\bibfnamefont {P.}~\bibnamefont
  {Nordlander}}, \ and\ \bibinfo {author} {\bibfnamefont {N.~J.}\ \bibnamefont
  {Halas}},\ }\href {\doibase 10.1021/acs.jpclett.7b00563} {\bibfield
  {journal} {\bibinfo  {journal} {J. Phys. Chem. Lett.}\ }\textbf {\bibinfo
  {volume} {8}},\ \bibinfo {pages} {2060} (\bibinfo {year} {2017})}\BibitemShut
  {NoStop}%
\bibitem [{\citenamefont {Sheldon}\ \emph {et~al.}(2014)\citenamefont
  {Sheldon}, \citenamefont {van~de Groep}, \citenamefont {Brown}, \citenamefont
  {Polman},\ and\ \citenamefont {Atwater}}]{Sheldon2014}%
  \BibitemOpen
  \bibfield  {author} {\bibinfo {author} {\bibfnamefont {M.~T.}\ \bibnamefont
  {Sheldon}}, \bibinfo {author} {\bibfnamefont {J.}~\bibnamefont {van~de
  Groep}}, \bibinfo {author} {\bibfnamefont {A.~M.}\ \bibnamefont {Brown}},
  \bibinfo {author} {\bibfnamefont {A.}~\bibnamefont {Polman}}, \ and\ \bibinfo
  {author} {\bibfnamefont {H.~A.}\ \bibnamefont {Atwater}},\ }\href {\doibase
  10.1126/science.1258405} {\bibfield  {journal} {\bibinfo  {journal}
  {Science}\ }\textbf {\bibinfo {volume} {346}},\ \bibinfo {pages} {828}
  (\bibinfo {year} {2014})}\BibitemShut {NoStop}%
\bibitem [{\citenamefont {Le~Ru}\ and\ \citenamefont
  {Etchegoin}(2008)}]{Le2008}%
  \BibitemOpen
  \bibfield  {author} {\bibinfo {author} {\bibfnamefont {E.}~\bibnamefont
  {Le~Ru}}\ and\ \bibinfo {author} {\bibfnamefont {P.}~\bibnamefont
  {Etchegoin}},\ }\href@noop {} {\emph {\bibinfo {title} {Principles of
  Surface-Enhanced Raman Spectroscopy: And Related Plasmonic Effects}}}\
  (\bibinfo  {publisher} {Elsevier, Oxford},\ \bibinfo {year}
  {2008})\BibitemShut {NoStop}%
\bibitem [{\citenamefont {Kabashin}\ \emph {et~al.}(2009)\citenamefont
  {Kabashin}, \citenamefont {Evans}, \citenamefont {Pastkovsky}, \citenamefont
  {Hendren}, \citenamefont {Wurtz}, \citenamefont {Atkinson}, \citenamefont
  {Pollard}, \citenamefont {Podolskiy},\ and\ \citenamefont
  {Zayats}}]{Kabashin2009}%
  \BibitemOpen
  \bibfield  {author} {\bibinfo {author} {\bibfnamefont {A.~V.}\ \bibnamefont
  {Kabashin}}, \bibinfo {author} {\bibfnamefont {P.}~\bibnamefont {Evans}},
  \bibinfo {author} {\bibfnamefont {S.}~\bibnamefont {Pastkovsky}}, \bibinfo
  {author} {\bibfnamefont {W.}~\bibnamefont {Hendren}}, \bibinfo {author}
  {\bibfnamefont {G.~A.}\ \bibnamefont {Wurtz}}, \bibinfo {author}
  {\bibfnamefont {R.}~\bibnamefont {Atkinson}}, \bibinfo {author}
  {\bibfnamefont {R.}~\bibnamefont {Pollard}}, \bibinfo {author} {\bibfnamefont
  {V.~A.}\ \bibnamefont {Podolskiy}}, \ and\ \bibinfo {author} {\bibfnamefont
  {A.~V.}\ \bibnamefont {Zayats}},\ }\href {https://doi.org/10.1038/nmat2546}
  {\bibfield  {journal} {\bibinfo  {journal} {Nat. Mater.}\ }\textbf {\bibinfo
  {volume} {8}},\ \bibinfo {pages} {867} (\bibinfo {year} {2009})}\BibitemShut
  {NoStop}%
\bibitem [{\citenamefont {Ayala-Orozco}\ \emph {et~al.}(2014)\citenamefont
  {Ayala-Orozco}, \citenamefont {Urban}, \citenamefont {Knight}, \citenamefont
  {Urban}, \citenamefont {Neumann}, \citenamefont {Bishnoi}, \citenamefont
  {Mukherjee}, \citenamefont {Goodman}, \citenamefont {Charron}, \citenamefont
  {Mitchell}, \citenamefont {Shea}, \citenamefont {Roy}, \citenamefont {Nanda},
  \citenamefont {Schiff}, \citenamefont {Halas},\ and\ \citenamefont
  {Joshi}}]{Ayala2014}%
  \BibitemOpen
  \bibfield  {author} {\bibinfo {author} {\bibfnamefont {C.}~\bibnamefont
  {Ayala-Orozco}}, \bibinfo {author} {\bibfnamefont {C.}~\bibnamefont {Urban}},
  \bibinfo {author} {\bibfnamefont {M.~W.}\ \bibnamefont {Knight}}, \bibinfo
  {author} {\bibfnamefont {A.~S.}\ \bibnamefont {Urban}}, \bibinfo {author}
  {\bibfnamefont {O.}~\bibnamefont {Neumann}}, \bibinfo {author} {\bibfnamefont
  {S.~W.}\ \bibnamefont {Bishnoi}}, \bibinfo {author} {\bibfnamefont
  {S.}~\bibnamefont {Mukherjee}}, \bibinfo {author} {\bibfnamefont {A.~M.}\
  \bibnamefont {Goodman}}, \bibinfo {author} {\bibfnamefont {H.}~\bibnamefont
  {Charron}}, \bibinfo {author} {\bibfnamefont {T.}~\bibnamefont {Mitchell}},
  \bibinfo {author} {\bibfnamefont {M.}~\bibnamefont {Shea}}, \bibinfo {author}
  {\bibfnamefont {R.}~\bibnamefont {Roy}}, \bibinfo {author} {\bibfnamefont
  {S.}~\bibnamefont {Nanda}}, \bibinfo {author} {\bibfnamefont
  {R.}~\bibnamefont {Schiff}}, \bibinfo {author} {\bibfnamefont {N.~J.}\
  \bibnamefont {Halas}}, \ and\ \bibinfo {author} {\bibfnamefont
  {A.}~\bibnamefont {Joshi}},\ }\href {\doibase 10.1021/nn501871d} {\bibfield
  {journal} {\bibinfo  {journal} {ACS Nano}\ }\textbf {\bibinfo {volume} {8}},\
  \bibinfo {pages} {6372} (\bibinfo {year} {2014})}\BibitemShut {NoStop}%
\bibitem [{\citenamefont {Krausz}\ and\ \citenamefont
  {Stockman}(2014)}]{Krausz2014}%
  \BibitemOpen
  \bibfield  {author} {\bibinfo {author} {\bibfnamefont {F.}~\bibnamefont
  {Krausz}}\ and\ \bibinfo {author} {\bibfnamefont {M.~I.}\ \bibnamefont
  {Stockman}},\ }\href {https://doi.org/10.1038/nphoton.2014.28} {\bibfield
  {journal} {\bibinfo  {journal} {Nat. Photon.}\ }\textbf {\bibinfo {volume}
  {8}},\ \bibinfo {pages} {205} (\bibinfo {year} {2014})}\BibitemShut {NoStop}%
\bibitem [{Nat()}]{NatFELcollect2019}%
  \BibitemOpen
  \href@noop {} {}\bibinfo {note} {`Free-Electron Lasers'. A collection of
  recent articles on FEL generation and characterization and their application
  in fundamental studies of light-matter interaction. Nature Photonics
  Collection (January 23, 2019)}\BibitemShut {NoStop}%
\bibitem [{\citenamefont {Merzbacher}(1998)}]{Merzbacher1998}%
  \BibitemOpen
  \bibfield  {author} {\bibinfo {author} {\bibfnamefont {E.}~\bibnamefont
  {Merzbacher}},\ }\href@noop {} {\emph {\bibinfo {title} {{Quantum
  Mechanics}}}},\ \bibinfo {edition} {3rd}\ ed.\ (\bibinfo  {publisher}
  {Wiley},\ \bibinfo {year} {1998})\ pp.\ \bibinfo {pages} {115, 315ff, 491,
  496}\BibitemShut {NoStop}%
\bibitem [{\citenamefont {Obreshkov}\ and\ \citenamefont
  {Thumm}(2006)}]{Obreshkov2006}%
  \BibitemOpen
  \bibfield  {author} {\bibinfo {author} {\bibfnamefont {B.}~\bibnamefont
  {Obreshkov}}\ and\ \bibinfo {author} {\bibfnamefont {U.}~\bibnamefont
  {Thumm}},\ }\href {\doibase 10.1103/PhysRevA.74.012901} {\bibfield  {journal}
  {\bibinfo  {journal} {Phys. Rev. A}\ }\textbf {\bibinfo {volume} {74}},\
  \bibinfo {pages} {012901} (\bibinfo {year} {2006})}\BibitemShut {NoStop}%
\bibitem [{\citenamefont {Liao}\ and\ \citenamefont
  {Thumm}(2014{\natexlab{a}})}]{Liao2014PRL}%
  \BibitemOpen
  \bibfield  {author} {\bibinfo {author} {\bibfnamefont {Q.}~\bibnamefont
  {Liao}}\ and\ \bibinfo {author} {\bibfnamefont {U.}~\bibnamefont {Thumm}},\
  }\href {\doibase 10.1103/PhysRevLett.112.023602} {\bibfield  {journal}
  {\bibinfo  {journal} {Phys. Rev. Lett.}\ }\textbf {\bibinfo {volume} {112}},\
  \bibinfo {pages} {023602} (\bibinfo {year} {2014}{\natexlab{a}})}\BibitemShut
  {NoStop}%
\bibitem [{\citenamefont {Liao}\ and\ \citenamefont
  {Thumm}(2014{\natexlab{b}})}]{Liao2014}%
  \BibitemOpen
  \bibfield  {author} {\bibinfo {author} {\bibfnamefont {Q.}~\bibnamefont
  {Liao}}\ and\ \bibinfo {author} {\bibfnamefont {U.}~\bibnamefont {Thumm}},\
  }\href {\doibase 10.1103/PhysRevA.89.033849} {\bibfield  {journal} {\bibinfo
  {journal} {Phys. Rev. A}\ }\textbf {\bibinfo {volume} {89}},\ \bibinfo
  {pages} {033849} (\bibinfo {year} {2014}{\natexlab{b}})}\BibitemShut
  {NoStop}%
\bibitem [{\citenamefont {Neppl}\ \emph
  {et~al.}(2015{\natexlab{b}})\citenamefont {Neppl}, \citenamefont
  {Ernstorfer}, \citenamefont {Cavalieri}, \citenamefont {Lemell},
  \citenamefont {Wachter}, \citenamefont {Magerl}, \citenamefont
  {Bothschafter}, \citenamefont {Jobst}, \citenamefont {Hofstetter},
  \citenamefont {Kleineberg}, \citenamefont {Barth}, \citenamefont {Menzel},
  \citenamefont {Burgd{\"o}rfer}, \citenamefont {Feulner}, \citenamefont
  {Krausz},\ and\ \citenamefont {Kienberger}}]{Neppl2015Nature}%
  \BibitemOpen
  \bibfield  {author} {\bibinfo {author} {\bibfnamefont {S.}~\bibnamefont
  {Neppl}}, \bibinfo {author} {\bibfnamefont {R.}~\bibnamefont {Ernstorfer}},
  \bibinfo {author} {\bibfnamefont {A.~L.}\ \bibnamefont {Cavalieri}}, \bibinfo
  {author} {\bibfnamefont {C.}~\bibnamefont {Lemell}}, \bibinfo {author}
  {\bibfnamefont {G.}~\bibnamefont {Wachter}}, \bibinfo {author} {\bibfnamefont
  {E.}~\bibnamefont {Magerl}}, \bibinfo {author} {\bibfnamefont {E.~M.}\
  \bibnamefont {Bothschafter}}, \bibinfo {author} {\bibfnamefont
  {M.}~\bibnamefont {Jobst}}, \bibinfo {author} {\bibfnamefont
  {M.}~\bibnamefont {Hofstetter}}, \bibinfo {author} {\bibfnamefont
  {U.}~\bibnamefont {Kleineberg}}, \bibinfo {author} {\bibfnamefont {J.~V.}\
  \bibnamefont {Barth}}, \bibinfo {author} {\bibfnamefont {D.}~\bibnamefont
  {Menzel}}, \bibinfo {author} {\bibfnamefont {J.}~\bibnamefont
  {Burgd{\"o}rfer}}, \bibinfo {author} {\bibfnamefont {P.}~\bibnamefont
  {Feulner}}, \bibinfo {author} {\bibfnamefont {F.}~\bibnamefont {Krausz}}, \
  and\ \bibinfo {author} {\bibfnamefont {R.}~\bibnamefont {Kienberger}},\
  }\href {https://doi.org/10.1038/nature14094} {\bibfield  {journal} {\bibinfo
  {journal} {Nature}\ }\textbf {\bibinfo {volume} {517}},\ \bibinfo {pages}
  {342} (\bibinfo {year} {2015}{\natexlab{b}})}\BibitemShut {NoStop}%
\bibitem [{\citenamefont {Zhang}\ and\ \citenamefont
  {Thumm}(2010)}]{Zhang2010}%
  \BibitemOpen
  \bibfield  {author} {\bibinfo {author} {\bibfnamefont {C.-H.}\ \bibnamefont
  {Zhang}}\ and\ \bibinfo {author} {\bibfnamefont {U.}~\bibnamefont {Thumm}},\
  }\href {\doibase 10.1103/PhysRevA.82.043405} {\bibfield  {journal} {\bibinfo
  {journal} {Phys. Rev. A}\ }\textbf {\bibinfo {volume} {82}},\ \bibinfo
  {pages} {043405} (\bibinfo {year} {2010})}\BibitemShut {NoStop}%
\bibitem [{\citenamefont {Wolkow}(1935)}]{Volkov1935}%
  \BibitemOpen
  \bibfield  {author} {\bibinfo {author} {\bibfnamefont {D.~M.}\ \bibnamefont
  {Wolkow}},\ }\href {\doibase 10.1007/BF01331022} {\bibfield  {journal}
  {\bibinfo  {journal} {Zeitschrift f{\"u}r Physik}\ }\textbf {\bibinfo
  {volume} {94}},\ \bibinfo {pages} {250} (\bibinfo {year} {1935})}\BibitemShut
  {NoStop}%
\bibitem [{\citenamefont {Zhang}\ and\ \citenamefont
  {Thumm}(2009)}]{Zhang2009PRL}%
  \BibitemOpen
  \bibfield  {author} {\bibinfo {author} {\bibfnamefont {C.-H.}\ \bibnamefont
  {Zhang}}\ and\ \bibinfo {author} {\bibfnamefont {U.}~\bibnamefont {Thumm}},\
  }\href {\doibase 10.1103/PhysRevLett.102.123601} {\bibfield  {journal}
  {\bibinfo  {journal} {Phys. Rev. Lett.}\ }\textbf {\bibinfo {volume} {102}},\
  \bibinfo {pages} {123601} (\bibinfo {year} {2009})}\BibitemShut {NoStop}%
\bibitem [{\citenamefont {Boiron}\ and\ \citenamefont
  {Lombardi}(1998)}]{Boiron1998}%
  \BibitemOpen
  \bibfield  {author} {\bibinfo {author} {\bibfnamefont {M.}~\bibnamefont
  {Boiron}}\ and\ \bibinfo {author} {\bibfnamefont {M.}~\bibnamefont
  {Lombardi}},\ }\href {\doibase 10.1063/1.475743} {\bibfield  {journal}
  {\bibinfo  {journal} {J. Chem. Phys.}\ }\textbf {\bibinfo {volume} {108}},\
  \bibinfo {pages} {3431} (\bibinfo {year} {1998})}\BibitemShut {NoStop}%
\bibitem [{\citenamefont {Goldfarb}\ \emph {et~al.}(2008)\citenamefont
  {Goldfarb}, \citenamefont {Schiff},\ and\ \citenamefont
  {Tannor}}]{Goldfarb2008}%
  \BibitemOpen
  \bibfield  {author} {\bibinfo {author} {\bibfnamefont {Y.}~\bibnamefont
  {Goldfarb}}, \bibinfo {author} {\bibfnamefont {J.}~\bibnamefont {Schiff}}, \
  and\ \bibinfo {author} {\bibfnamefont {D.~J.}\ \bibnamefont {Tannor}},\
  }\href {\doibase 10.1063/1.2907336} {\bibfield  {journal} {\bibinfo
  {journal} {J. Chem. Phys.}\ }\textbf {\bibinfo {volume} {128}},\ \bibinfo
  {pages} {164114} (\bibinfo {year} {2008})}\BibitemShut {NoStop}%
\bibitem [{\citenamefont {Goldstein}\ \emph {et~al.}(2001)\citenamefont
  {Goldstein}, \citenamefont {Poole},\ and\ \citenamefont {Safko}}]{Goldstein}%
  \BibitemOpen
  \bibfield  {author} {\bibinfo {author} {\bibfnamefont {H.}~\bibnamefont
  {Goldstein}}, \bibinfo {author} {\bibfnamefont {C.}~\bibnamefont {Poole}}, \
  and\ \bibinfo {author} {\bibfnamefont {J.}~\bibnamefont {Safko}},\
  }\href@noop {} {\emph {\bibinfo {title} {Classical mechanics}}},\ \bibinfo
  {edition} {3rd}\ ed.\ (\bibinfo  {publisher} {Addison-wesley},\ \bibinfo
  {year} {2001})\ pp.\ \bibinfo {pages} {22--23,342,433}\BibitemShut {NoStop}%
\bibitem [{\citenamefont {Broyden}(1965)}]{Broyden}%
  \BibitemOpen
  \bibfield  {author} {\bibinfo {author} {\bibfnamefont {C.~G.}\ \bibnamefont
  {Broyden}},\ }\href@noop {} {\bibfield  {journal} {\bibinfo  {journal}
  {Mathematics of Computation}\ }\textbf {\bibinfo {volume} {19}},\ \bibinfo
  {pages} {577} (\bibinfo {year} {1965})}\BibitemShut {NoStop}%
\bibitem [{\citenamefont {Atkinson}(2008)}]{Atkinson}%
  \BibitemOpen
  \bibfield  {author} {\bibinfo {author} {\bibfnamefont {K.~E.}\ \bibnamefont
  {Atkinson}},\ }\href@noop {} {\emph {\bibinfo {title} {An introduction to
  numerical analysis}}}\ (\bibinfo  {publisher} {John Wiley \& Sons},\ \bibinfo
  {year} {2008})\BibitemShut {NoStop}%
\bibitem [{\citenamefont {Batchelor}(1967)}]{Batchelor}%
  \BibitemOpen
  \bibfield  {author} {\bibinfo {author} {\bibfnamefont {G.~K.}\ \bibnamefont
  {Batchelor}},\ }\href@noop {} {\emph {\bibinfo {title} {{An introduction to
  fluid dynamics}}}}\ (\bibinfo  {publisher} {Cambridge University Press},\
  \bibinfo {year} {1967})\ p.~\bibinfo {pages} {74}\BibitemShut {NoStop}%
\bibitem [{\citenamefont {Landau}\ and\ \citenamefont
  {Lifshitz}(1977)}]{Landau}%
  \BibitemOpen
  \bibfield  {author} {\bibinfo {author} {\bibfnamefont {L.~D.}\ \bibnamefont
  {Landau}}\ and\ \bibinfo {author} {\bibfnamefont {E.~M.}\ \bibnamefont
  {Lifshitz}},\ }\href@noop {} {\emph {\bibinfo {title} {{Quantum mechanics:
  non-relativistic theory}}}},\ \bibinfo {edition} {3rd}\ ed.\ (\bibinfo
  {publisher} {Elsevier},\ \bibinfo {year} {1977})\ p.\ \bibinfo {pages}
  {570}\BibitemShut {NoStop}%
\bibitem [{\citenamefont {Rosenberg}\ and\ \citenamefont
  {Zhou}(1993)}]{Rosenberg1993}%
  \BibitemOpen
  \bibfield  {author} {\bibinfo {author} {\bibfnamefont {L.}~\bibnamefont
  {Rosenberg}}\ and\ \bibinfo {author} {\bibfnamefont {F.}~\bibnamefont
  {Zhou}},\ }\href {\doibase 10.1103/PhysRevA.47.2146} {\bibfield  {journal}
  {\bibinfo  {journal} {Phys. Rev. A}\ }\textbf {\bibinfo {volume} {47}},\
  \bibinfo {pages} {2146} (\bibinfo {year} {1993})}\BibitemShut {NoStop}%
\bibitem [{\citenamefont {Reiss}\ and\ \citenamefont
  {Krainov}(1994)}]{Reiss1994CV}%
  \BibitemOpen
  \bibfield  {author} {\bibinfo {author} {\bibfnamefont {H.~R.}\ \bibnamefont
  {Reiss}}\ and\ \bibinfo {author} {\bibfnamefont {V.~P.}\ \bibnamefont
  {Krainov}},\ }\href {\doibase 10.1103/PhysRevA.50.R910} {\bibfield  {journal}
  {\bibinfo  {journal} {Phys. Rev. A}\ }\textbf {\bibinfo {volume} {50}},\
  \bibinfo {pages} {R910} (\bibinfo {year} {1994})}\BibitemShut {NoStop}%
\bibitem [{\citenamefont {Macri}\ \emph {et~al.}(2003)\citenamefont {Macri},
  \citenamefont {Miraglia},\ and\ \citenamefont {Gravielle}}]{Macri2003CV}%
  \BibitemOpen
  \bibfield  {author} {\bibinfo {author} {\bibfnamefont {P.~A.}\ \bibnamefont
  {Macri}}, \bibinfo {author} {\bibfnamefont {J.~E.}\ \bibnamefont {Miraglia}},
  \ and\ \bibinfo {author} {\bibfnamefont {M.~S.}\ \bibnamefont {Gravielle}},\
  }\href {\doibase 10.1364/JOSAB.20.001801} {\bibfield  {journal} {\bibinfo
  {journal} {J. Opt. Soc. Am. B}\ }\textbf {\bibinfo {volume} {20}},\ \bibinfo
  {pages} {1801} (\bibinfo {year} {2003})}\BibitemShut {NoStop}%
\bibitem [{\citenamefont {Dubois}\ \emph {et~al.}(2019)\citenamefont {Dubois},
  \citenamefont {Berman}, \citenamefont {Chandre},\ and\ \citenamefont
  {Uzer}}]{Uzer2019}%
  \BibitemOpen
  \bibfield  {author} {\bibinfo {author} {\bibfnamefont {J.}~\bibnamefont
  {Dubois}}, \bibinfo {author} {\bibfnamefont {S.~A.}\ \bibnamefont {Berman}},
  \bibinfo {author} {\bibfnamefont {C.}~\bibnamefont {Chandre}}, \ and\
  \bibinfo {author} {\bibfnamefont {T.}~\bibnamefont {Uzer}},\ }\href {\doibase
  10.1103/PhysRevA.99.053405} {\bibfield  {journal} {\bibinfo  {journal} {Phys.
  Rev. A}\ }\textbf {\bibinfo {volume} {99}},\ \bibinfo {pages} {053405}
  (\bibinfo {year} {2019})}\BibitemShut {NoStop}%
\bibitem [{Sup()}]{SupplMat}%
  \BibitemOpen
  \href@noop {} {}\bibinfo {note} {See Supplemental Material at {\red [url]}
  for animations of the (a) wavefunction comparison and (b)
  Mie-theory-calculated and reconstructed electric near-field distributions for
  Au nanospheres.}\BibitemShut {Stop}%
\bibitem [{\citenamefont {Patchkovskii}\ and\ \citenamefont
  {Muller}(2016)}]{SCID-TDSE}%
  \BibitemOpen
  \bibfield  {author} {\bibinfo {author} {\bibfnamefont {S.}~\bibnamefont
  {Patchkovskii}}\ and\ \bibinfo {author} {\bibfnamefont {H.}~\bibnamefont
  {Muller}},\ }\href {\doibase https://doi.org/10.1016/j.cpc.2015.10.014}
  {\bibfield  {journal} {\bibinfo  {journal} {Comp. Phys. Commun.}\ }\textbf
  {\bibinfo {volume} {199}},\ \bibinfo {pages} {153} (\bibinfo {year}
  {2016})}\BibitemShut {NoStop}%
\bibitem [{\citenamefont {Mie}(1908)}]{Mie1908}%
  \BibitemOpen
  \bibfield  {author} {\bibinfo {author} {\bibfnamefont {G.}~\bibnamefont
  {Mie}},\ }\href {http://dx.doi.org/10.1002/andp.19083300302} {\bibfield
  {journal} {\bibinfo  {journal} {Ann. Phys. (Berlin, Ger.)}\ }\textbf
  {\bibinfo {volume} {330}},\ \bibinfo {pages} {377} (\bibinfo {year}
  {1908})}\BibitemShut {NoStop}%
\end{thebibliography}%

\end{document}